\documentclass[11pt]{llncs}
\usepackage{amssymb,amsmath,enumerate,graphicx}

\usepackage{tikz}
\usetikzlibrary{snakes,shapes}
\tikzstyle{hyb}=[rectangle,draw,minimum size=2mm]
\tikzstyle{tre}=[circle,draw,minimum size=2mm]
\tikzstyle{tri}=[regular polygon,regular polygon sides=5,draw,minimum size=2mm]
\newcommand{\etq}[1]{%
\draw (#1) node {\tiny $#1$};
}

\newcommand{\pathgr}{\!\rightsquigarrow\!{}}
\newcommand{\cC}{\mathcal{C}}
\newcommand{\cIC}{\mathcal{C}_I}
\newcommand{\cIT}{\theta_I}

\renewcommand{\leq}{\leqslant}
\renewcommand{\geq}{\geqslant}

\newcommand{\RR}{\mathbb{R}}
\newcommand{\ovG}{\overline{G}}

\begin{document}

\title{Comparison of Galled Trees}

\author{Gabriel Cardona\inst{1}\and Merc\`e Llabr\'es\inst{1}\and Francesc Rossell\'o\inst{1} \and
Gabriel Valiente\inst{2}} 
\institute{Department of Mathematics and Computer Science, University
of the Balearic Islands, E-07122 Palma de Mallorca,
\texttt{\{gabriel.cardona,merce.llabres,cesc.rossello\}@uib.es} \and
Algorithms, Bioinformatics, Complexity and Formal Methods Research
Group, Technical University of Catalonia, E-08034 Barcelona,
\texttt{valiente@lsi.upc.edu} }
\maketitle

\begin{abstract}
Galled trees, directed acyclic graphs that model evolutionary
histories with isolated hybridization events, have become very popular
due to both their biological significance and the existence of
polynomial time algorithms for their reconstruction. In this
paper we establish to which extent several distance measures for the
comparison of evolutionary networks are metrics for galled trees, and
hence when they can be safely used to evaluate galled tree
reconstruction methods.
\end{abstract}

\section{Introduction}

The study of phylogenetic networks as a model of reticulate evolution
began with the representation of conflicting phylogenetic signals as
an implicit \emph{splits
nework}~\cite{bandelt.dress:1992,huson.bryant:2006}, but it was soon
realized that internal nodes in a splits network did not have any
direct interpretation in evolutionary terms.  Attention turned then to
the study of explicit evolutionary networks, in which the internal
nodes have a direct interpretation as reticulate evolutionary events
such as recombination, hybridization, or lateral gene transfer. Unfortunately, 
the hardness of reconstructing an evolutionary network with as few
recombination events as possible for a set of sequences, under the
assumption of no repeated or back mutations, was soon
established~\cite{bordewich.sample:2004,wang.ea:2000,wang.ea:2001}.

However, when the conflicting phylogenetic signals show a particular structure,
such that the \emph{conflict graph} of the set of sequences is
\emph{biconvex}, the evolutionary network with the smallest
possible number of recombination events
is unique, it can be reconstructed in polynomial time and it is a \emph{galled tree},
an evolutionary network with hybrid nodes of in-degree 2 (because they correspond to explicit recombinations) and disjoint reticulation cycles~\cite{gusfield.ea:csb:2003}.
Galled trees are also relevant from a biological point of view because, as Gusfield \textsl{et al} point out in \textsl{loc. cit.}, reticulation events tend to be isolated, yielding to disjoint reticulation cycles,  if the level of recombination is moderate, or
if most of the observable recombinations are recent.
Actually, several slightly different notions of galled tree have been introduced so far in the literature, depending on the degree of disjointness of their reticulation cycles.
The original galled trees \cite{gusfield.ea:csb:2003} have node-disjoint reticulation cycles, while the \emph{nested networks with nesting level 1}~\cite{JanssonSung2004a,JanssonSung2008} (dubbed, for simplicity, \emph{1-nested networks} in this paper) have arc-disjoint  reticulation cycles. Between both notions lie the level-1 networks~\cite{CJSS2005,jansson.sung:2006}, without biconnected components with more than one hybrid node.  We have studied the relationships among these types of networks~\cite{rv:galled09}: see Section~\ref{sec:gt} below.

Now, various algorithms are known for reconstructing galled trees from
either
sequences~\cite{gusfield.ea:2007,gusfield.ea:csb:2003,gusfield.ea:informs:2004,gusfield.ea:2004},
trees~\cite{nakhleh.ea:2005}, distances~\cite{chan.ea:2006},
splits~\cite{huson.kloepper:2007}, or triplets~\cite{jansson.ea:2006},
and metrics provide a safe way to assess phylogenetic reconstruction
methods~\cite{miyamoto.cracraft:1991,nakhleh.ea:03psb}. A few
polynomial time computable metrics, like the path multiplicity or
$\mu$-distance \cite{cardona.ea:07b} or Nakhleh's metric $m$ for reduced networks \cite{nakhleh:2009}, 
 are known for \emph{tree child} evolutionary networks~\cite{cardona.ea:07b}, which include galled trees \cite{rv:galled09}. But most distance measures introduced so far were only known to be metrics on \emph{time consistent}~\cite{baroni.ea:2006}
 tree-child phylogenetic
networks, including the Robinson-Foulds
distance~\cite{baroni.ea:ac04,cardona.ea:07a,cardona.ea:tcbb:comparison.1:2008}, the tripartitions
distance~\cite{cardona.ea:07a,moret.ea:2004},
the nodal and splitted nodal
distances~\cite{nodalnetworks,cardona.ea:tcbb:comparison.2:2008}, and
the triplets distance~\cite{cardona.ea:tcbb:comparison.2:2008}. 
Since galled trees need not be time consistent, it was not known whether these distance measures define metrics for galled trees. On the other hand,  Nakhleh gave
in his PhD Thesis \cite{nakhleh:phd04} two metrics for time consistent galled trees
(based on splits and subtrees), but they are not metrics for arbitrary galled
trees \cite{cardona.ea:07b}. Recent simulation studies using the
coalescent model with recombination show that only a small fraction of
the simulated galled trees are time consistent~\cite{arenas.ea:2008}.

In this paper, we study which of the aforementioned metrics for tree-child time consistent phylogenetic networks are also metrics for galled trees, under the various notions of the latter. We show that the Robinson-Foulds distance is only a metric in the binary case (in which the original galled trees, the level-1 networks and the 1-nested networks are the same objects); the tripartitions distance is a metric for 1-nested networks  without any restriction on the degrees of their nodes (besides the general restriction that hybrid nodes have in-degree 2); and the splitted nodal distance is a metric in the \emph{semibinary} (hybrid nodes of in-degree 2 and out-degree 1) case, in which the 1-nested and level-1 conditions define the same objects, but they are strictly weaker than the node-disjoint reticulation cycles condition). On the other hand, neither the nodal distance nor the triplets distance are metrics even for the most restrictive case of binary galled trees.

\section{Preliminaries}
\label{sec:prel}

Given a set $S$, a \emph{$S$-rDAG} is a rooted directed acyclic graph
with its leaves bijectively labeled in $S$.

A \emph{tree node} of a $S$-rDAG $N=(V,E)$ is a node of in-degree at
most 1, and a \emph{hybrid node} is a node of in-degree at least 2.  A
\emph{tree arc} (respectively, a \emph{hybridization arc}) is an arc
with head a tree node (respectively, a hybrid node).  A node $v\in V$
is a \emph{child} of $u\in V$ if $(u,v)\in E$; we also say in this
case that $u$ is a \emph{parent} of $v$.  Two nodes are \emph{sibling}
when they have a common parent.

We denote by $u\pathgr v$ any path in $N$ with origin $u$ and end $v$.
Whenever there exists a path $u\pathgr v$, we shall say that $v$ is a
\emph{descendant} of $u$ and also that $u$ is an \emph{ancestor} of
$v$.  The \emph{length} of a path is its number of arcs, and the
\emph{distance} from a node $u$ to a descendant $v$ of it is the
length of a shortest path $u\pathgr v$.

A node $v$ is a \emph{strict} descendant of a node $u$ in $N$ when
every path from the root of $N$ to $v$ contains the node $u$; thus,
$v$ is a \emph{non-strict} descendant of $u$ when it is a descendant
of $u$, but there exist paths from the root to $v$ that do not contain
$u$.  The following straightforward result, which is Lemma 1 in
\cite{cardona.ea:tcbb:comparison.1:2008}, will be used often, usually
without any further notice.

\begin{lemma}
\label{lem:strict->anc}
Every strict ancestor of a node $v$ is connected by a path with every
ancestor of $v$. \qed
\end{lemma}

A \emph{tree path} is a path consisting only of tree arcs, and a
node $v$ is a \emph{tree descendant} of a node $u$ when there is a
tree path $u\pathgr v$.  The following result summarizes Lemma 3 and
Corollary 4 in \cite{cardona.ea:07a}, and it will also be used many
times in this paper without
any further notice.

\begin{lemma}\label{lem:unicity-path}
Let $u\pathgr v$ be a tree path in a $S$-rDAG. 
\begin{enumerate}[(1)]

\item Every other path $w\pathgr v$ ending in $v$  either is
contained in $u\pathgr v$ or contains $u\pathgr v$.  In particular,
if $w$ is a descendant of $u$ and there exists a path $w\pathgr v$,
then this path is contained in the tree path $u\pathgr v$.

\item The tree path $u\pathgr v$ is the unique path from $u$ to $v$.

\item The node $v$ is a strict descendant of $u$. \qed
\end{enumerate}
\end{lemma}

Two paths in a $S$-rDAG are \emph{internally disjoint} when they have
disjoint sets of intermediate nodes.  A \emph{reticulation cycle} for
a hybrid node $h$ is a pair of internally disjoint paths ending in $h$
and with the same origin.  Each one of the paths forming a
reticulation cycle for $h$ is called generically a \emph{merge path},
their common origin is called the \emph{split node} of the
reticulation cycle, and the hybrid node $h$, its \emph{end}.  The
\emph{intermediate nodes} of a reticulation cycle are the intermediate
nodes of the merge paths forming it.

A subgraph of an undirected graph is \emph{biconnected} when it is
connected and it remains connected if we remove any node and all edges
incident to it.  A subgraph of a $S$-rDAG $N$ is said to be
\emph{biconnected} when it is so in the undirected graph associated to
$N$.

\section{1-nested networks}
\label{sec:gt}

In the rest of this paper, by a \emph{hybridization network} on a set
$S$ we understand a $S$-rDAG without out-degree 1 tree nodes and with
all its hybrid nodes of in-degree 2.  We shall also use the term
\emph{hybridization network with $n$ leaves} to refer to a
hybridization network on a set $S$ with $n$ elements.  A
\emph{phylogenetic tree} is a hybridization network without hybrid
nodes.

We shall say that a hybridization network is \emph{semibinary} when
its hybrid nodes have out-degree 1, and that it is \emph{binary} when
it is semibinary and its internal tree nodes have out-degree 2.

A hybridization network is:
\begin{itemize}

\item a \emph{galled tree}, when every pair of reticulation cycles
have disjoint sets of nodes \cite{gusfield.ea:csb:2003}.

\item \emph{1-nested}, when every pair of reticulation cycles have
disjoint sets of arcs: by \cite[Prop.~12]{rv:galled09}, this is
equivalent to the fact that every pair of reticulation cycles for
different hybrid nodes have disjoint sets of intermediate nodes, and
hence it corresponds to the notion of \emph{nested (hybridization)
network with nesting depth 1} \cite{JanssonSung2004a,JanssonSung2008}.

\item \emph{level-1}, when no biconnected component contains more than
1 hybrid node \cite{CJSS2005,jansson.sung:2006}.
\end{itemize}
To simplify the language, from now on we shall write simply
\emph{1-nested network} to mean a 1-nested hybridization network.  The
following two results summarize the main results on 1-nested networks
proved in \cite{rv:galled09}.

\begin{lemma}\label{1n no ih}
In a 1-nested network, every hybrid node is the end of exactly one
reticulation cycle, and all the intermediate nodes of this
reticulation cycle are of tree type.  \qed
\end{lemma}

\begin{theorem}\label{resum}
\begin{enumerate}[(a)]

\item Every 1-nested network is \emph{tree-child}, in the sense that
every internal node has a child of tree type.

\item For general hybridization networks,
$$
\mbox{galled tree}\Longrightarrow \mbox{level-1} \Longrightarrow
\mbox{1-nested},
$$
and these implications are strict.

\item For semibinary hybridization networks,
$$
\mbox{galled tree}\Longrightarrow \mbox{level-1} \Longleftrightarrow
\mbox{1-nested},
$$
and the first implication is strict.

\item For binary hybridization networks,
$$
\mbox{galled tree} \Longleftrightarrow \mbox{level-1}
\Longleftrightarrow \mbox{1-nested. \qed}
$$
\end{enumerate}
\end{theorem}

The fact that every 1-nested network is tree-child implies, by
\cite[Lem.~2]{cardona.ea:07a} the following result.

\begin{corollary}\label{lem:1n->tc}
Every node in a 1-nested   network has some tree
descendant leaf, and hence some strict descendant leaf. \qed
\end{corollary}

The following result lies at the basis of most of our proofs.

\begin{proposition}\label{prop:basic-2h1n}
Every 1-nested   network contains some internal tree node
with all its children tree leaves, or a hybrid node with all its
children tree leaves and such that all the intermediate nodes in its
reticulation cycle have all their children outside the reticulation
cycle tree leaves.
\end{proposition}

\begin{proof}
Let $N$ be a 1-nested network.  Let the \emph{galled-length} of a path
in $N$ be the number of reticulation cycles which the arcs of the path
belong to, and the \emph{galled-depth} of a node in $N$ the largest
galled-length of a path from the root to it.  Notice that the
galled-depth of a hybrid node is equal to the galled-depth of the
intermediate nodes of its reticulation cycle (because every arc in $N$
belongs at most to one reticulation cycle).

Assume that $N$ does not contain any internal tree node with all its
children tree leaves.  Let $h$ be a hybrid node of largest
galled-depth in $N$, and let $v$ denote either $h$ or any intermediate
node in the reticulation cycle $K$ for $h$.  It turns out that $v$ has
no hybrid descendant other than $h$, because any path from $v$ to any
other hybrid node $h'\neq h$ would contain arcs belonging to at least
one more reticulation cycle, making the galled-depth of $h'$ larger
than that of $v$.

Let $v'$ be any descendant of $v$ not belonging to $K$.  Then, $v'$ is
a tree node and all its descendants are tree nodes, and therefore,
since we assume that $N$ does not contain any internal tree node with
all its children tree leaves, we conclude that $v'$ is a tree leaf.
\end{proof}

\section{Reductions for 1-nested networks}
\label{sec:red-1n}

We introduce in this section a set of reductions for 1-nested
networks.  Each of these reductions, when applied to a 1-nested
network with $n$ leaves and $m$ nodes, produces a 1-nested network
with at most $n$ leaves and less than $m$ nodes, and given any
1-nested network with more than one leaf, it is always possible to
apply to it some of these reductions.  We shall also show that
suitable subsets of these reductions have similar properties for
binary and for semibinary 1-nested networks.  Similar sets of
reductions for other types of evolutionary networks have already been
published
\cite{cardona.ea:sbtstc:2008,cardona.ea:tcbb:comparison.2:2008}.
\smallskip

\begin{figure}[htb]
\centering
  \begin{tikzpicture}[thick,>=stealth,scale=0.3]
    \draw (0,2) node[tri] (xx) {};
    \draw (0,0) node[tri] (u) {}; 
    \draw (-1,-2) node[tre] (i) {}; \etq {i} 
    \draw (1,-2) node[tre] (j) {}; \etq {j} 
    \draw [->](xx)--(u); 
    \draw [->](u)--(i); 
    \draw [->](u)--(j);
  \end{tikzpicture}
  \qquad
               \begin{tikzpicture}[thick,>=stealth,scale=0.3]
              \draw(0,-2) node{\ };
              \draw(0,0) node  {$\Rightarrow$};
              \draw(0,2) node{\ };
 \end{tikzpicture}
  \qquad
  \begin{tikzpicture}[thick,>=stealth,scale=0.3]
    \draw (0,2) node[tri] (xx) {};
    \draw (0,0) node[tri] (u) {}; \draw (u) node {\tiny $i$};
    \draw (0,-2) node  {\ };  
    \draw [->](xx)--(u);
  \end{tikzpicture}

\caption{\label{fig:R-red} 
The $R_{i;j}$ reduction.}
\end{figure}
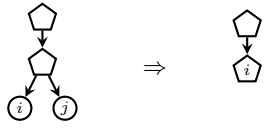

\noindent\textbf{\emph{The $R$ reductions.}} Let $N$ be a 1-nested
network with $n$ leaves, and let $u$ be an internal node whose
children are exactly the tree leaves $i$ and $j$.  The \emph{$R_{i;j}$
reduction} of $N$ is the network $R_{i;j}(N)$ obtained by removing the
leaves $i$ and $j$, together with their incoming arcs, and labeling
with $i$ their former common parent $u$, which has become now a leaf;
cf.~Fig.~\ref{fig:R-red}.\footnote{In graphical representations of
hybridization networks, we shall represent hybrid nodes by squares,
tree nodes by circles, and indeterminate (that is, that can be of tree
or hybrid type) nodes by pentagons.} It is clear that $R_{i;j}(N)$ is
a 1-nested network on $S\setminus\{j\}$, and it has 2 nodes less than
$N$.  \smallskip

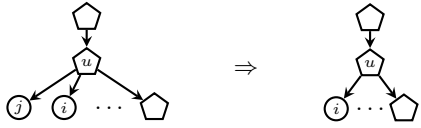
\begin{figure}[htb]
\centering
  \begin{tikzpicture}[thick,>=stealth,scale=0.3]
    \draw (0,2) node[tri] (xx) {};
    \draw (0,0) node[tri] (u) {}; \etq{u}
    \draw (-3,-2) node[tre] (j) {}; \etq {j}
    \draw (-1,-2) node[tre] (i) {};  \etq {i}
       \draw (1,-2) node  {$\ldots$}; 
    \draw (3,-2) node[tri] (ik) {}; 
 \draw [->](xx)--(u);
    \draw [->](u)--(i);
      \draw [->](u)--(j);
    \draw [->](u)--(ik);
  \end{tikzpicture}
  \qquad
               \begin{tikzpicture}[thick,>=stealth,scale=0.3]
              \draw(0,-2) node{\ };
              \draw(0,0) node  {$\Rightarrow$};
              \draw(0,2) node{\ };
 \end{tikzpicture}
  \qquad
  \begin{tikzpicture}[thick,>=stealth,scale=0.3]
     \draw (0,2) node[tri] (xx) {};
    \draw (0,0) node[tri] (u) {}; \etq{u}
    \draw (-1.5,-2) node[tre] (i) {};  \etq {i}
       \draw (0,-2) node  {$\ldots$}; 
    \draw (1.5,-2) node[tri] (ik) {}; 
 \draw [->](xx)--(u);
  \draw [->](u)--(i);
    \draw [->](u)--(ik);
  \end{tikzpicture}

\caption{\label{fig:T-red-1} 
The $T_{i;j}$ reduction.}
\end{figure}

\noindent\textbf{\emph{The $T$ reductions.}} Let $N$ be a 1-nested
network with $n$ leaves, and let $u$ be an internal node with two tree
leaf children $i,j$ and at least some other child.  The
\emph{$T_{i;j}$ reduction} of $N$ is the network $T_{i;j}(N)$ obtained
by removing the leaf $j$ together with its incoming arc;
cf.~Fig.~\ref{fig:T-red-1}.  It is clear that $T_{i;j}(N)$ is a
1-nested network on $S\setminus\{j\}$ with 1 node less than $N$.
\smallskip

\noindent\textbf{\emph{The $G$ reductions.}} Let $N$ be a 1-nested
network with $n$ leaves.  Assume that $N$ contains a reticulation
cycle $K$ consisting of two merge paths $(u,v_1,\ldots,v_k,h)$ and
$(u,v'_1,\ldots,v'_{k'},h)$, with $k\geq k'$ ($k'$ can be 0, in which
case the corresponding merge path is simply the arc $(u,h)$, but then
$k>0$), such that
\begin{itemize}

\item the hybrid node $h$ has only one child, and it is the tree leaf
$i$;

\item each intermediate node of $K$ has only one child outside $K$,
and it is a tree leaf: the child outside $K$ of each $v_j$ is the leaf
$i_j$ and the child outside $K$ of each $v'_j$ is the leaf $i_j'$.
\end{itemize}
Notice that $u$ may have children outside $K$.

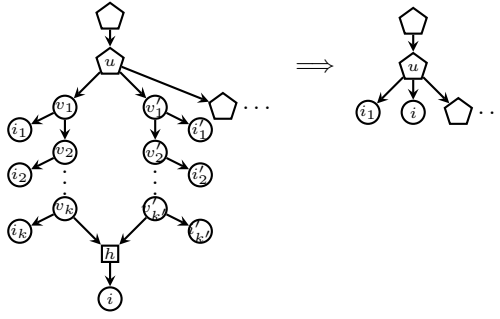
\begin{figure}[htb]
\centering
  \begin{tikzpicture}[thick,>=stealth,scale=0.3]
    \draw (0,2) node[tri] (xx) {};
    \draw (0,0) node[tri] (u) {}; \etq{u}
       \draw (5,-2) node[tri] (v) {};
  \draw (6.5,-2) node  {$\ldots$};
    \draw (-2,-2) node[tre] (v_1) {}; \etq{v_1}
    \draw (-2,-4) node[tre] (v_2) {}; \etq{v_2}
    \draw (-2,-5) node  {$\vdots$};  
    \draw (-2,-6.5) node[tre] (v_k) {}; \etq{v_k}
    \draw (0,-8.5) node[hyb] (h) {}; \etq{h}
 \draw (2,-2) node[tre] (v'_1) {}; \etq{v'_1}
    \draw (2,-4) node[tre] (v'_2) {}; \etq{v'_2}
    \draw (2,-5) node  {$\vdots$};  
    \draw (2,-6.5) node[tre] (v'_{k'}) {}; \etq{v'_{k'}}
   \draw (-4,-3) node[tre] (i_1) {}; \etq{i_1}
    \draw (-4,-5) node[tre] (i_2) {}; \etq{i_2}
    \draw (-4,-7.5) node[tre] (i_k) {}; \etq{i_k}
    \draw (0,-10.5) node[tre] (i) {}; \etq{i}
 \draw (4,-3) node[tre] (i'_1) {}; \etq{i'_1}
    \draw (4,-5) node[tre] (i'_2) {}; \etq{i'_2}
    \draw (4,-7.5) node[tre] (i'_{k'}) {}; \etq{i'_{k'}}
          \draw[->](u) to (v);
  \draw[->](xx)--(u);
    \draw[->](u) to (v_1);
        \draw[->](u) to (v'_1);
    \draw[->](v_1) to (i_1);
    \draw[->](v_1) to (v_2);
    \draw[->](v_2) to (i_2);
    \draw[->](v_k) to (i_k);
    \draw[->](v_k) to (h);    
    \draw[->](v'_1) to (i'_1);
    \draw[->](v'_1) to (v'_2);
     \draw[->](v'_2) to (i'_2);
   \draw[->](v'_{k'}) to (i'_{k'});
    \draw[->](v'_{k'}) to (h);    
    \draw[->](h) to (i);    
  \end{tikzpicture}
               \begin{tikzpicture}[thick,>=stealth,scale=0.3]
              \draw(0,2) node{\ };
              \draw(0,0) node  {$\Longrightarrow$};
              \draw(0,-10.5) node{\ };
 \end{tikzpicture}
  \begin{tikzpicture}[thick,>=stealth,scale=0.3]
     \draw (0,2) node[tri] (xx) {};
    \draw (0,0) node[tri] (u) {}; \etq{u}
    \draw (-2,-2) node[tre] (i_1) {}; \etq{i_1}
       \draw (0,-2) node[tre] (i) {}; \etq{i}
       \draw (2,-2) node[tri] (v) {};
       \draw (3.5,-2) node  {$\ldots$};
  \draw (0,-10.5) node  {\ };
   \draw[->](xx) to (u);
      \draw[->](u) to (i);
      \draw[->](u) to (i_1);
      \draw[->](u) to (v);

     \end{tikzpicture}

\caption{\label{fig:G-red} 
The $G_{i;i_1,\dots,i_k;i'_1,\dots,i'_{k'}}$ reduction.}
\end{figure}

The \emph{$G_{i;i_1,\dots,i_k;i'_1,\dots,i'_{k'}}$ reduction} of $N$
is the network $G_{i;i_1,\dots,i_k;i'_1,\dots,i'_{k'}}(N)$ obtained by
removing the nodes $v_1,\ldots,v_k,v_1',\ldots,v_k',h$ and the leaves
$i_1,\ldots,i_k,i'_1,\ldots,i'_{k'},i$, together with all their
incoming arcs, and then adding to the node $u$ two new tree leaf
children, labeled $i$ and $i_1$; cf.~Fig.~\ref{fig:G-red}.  Since we
remove a complete reticulation cycle and all descendants of its
intermediate nodes, and we replace it by two tree leaves, it is clear
that $G_{i;i_1,\dots,i_k;i'_1,\dots,i'_{k'}}(N)$ is a 1-nested network
on $S\setminus\{i_2,\dots,i_k,i'_1,\dots,i'_{k'}\}$ (and in
particular, if $k=1$ and $k'=0$, it has the same leaves as $N$) with
$2(k+k')$ nodes less than $N$.

\noindent\textbf{\emph{The $\ovG $ reductions.}} The
$\ovG_{i;i_1,\ldots,i_k; i'_1,\ldots,i'_{k'}}$ reduction is the same
as $G_{i;i_1,\dots,i_k;i'_1,\dots,i'_{k'}}$, except for the fact that
in order to apply the $\ovG_{i;i_1,\ldots,i_k; i'_1,\ldots,i'_{k'}}$
reduction, the hybrid node $h$ must be the leaf labeled $i$, instead
of the leaf's parent: see~Fig.~\ref{fig:G-red-1}.  Then,
$\ovG_{i;i_1,\dots,i_k;i'_1,\dots,i'_{k'}}(N)$ is a 1-nested network
on $S\setminus\{i_2,\dots,i_k,i'_1,\dots,i'_{k'}\}$ and it has
$2(k+k')-1$ nodes less than $N$.

\begin{figure}[htb]
\centering
  \begin{tikzpicture}[thick,>=stealth,scale=0.3]
    \draw (0,2) node[tri] (xx) {};
    \draw (0,0) node[tri] (u) {}; \etq{u}
     \draw (5,-2) node[tri] (v) {};
  \draw (6.5,-2) node  {$\ldots$};
    \draw (-2,-2) node[tre] (v_1) {}; \etq{v_1}
    \draw (-2,-4) node[tre] (v_2) {}; \etq{v_2}
    \draw (-2,-5) node  {$\vdots$};  
    \draw (-2,-6.5) node[tre] (v_k) {}; \etq{v_k}
    \draw (0,-8.5) node[hyb] (i) {}; \etq{i}
 \draw (2,-2) node[tre] (v'_1) {}; \etq{v'_1}
    \draw (2,-4) node[tre] (v'_2) {}; \etq{v'_2}
    \draw (2,-5) node  {$\vdots$};  
    \draw (2,-6.5) node[tre] (v'_{k'}) {}; \etq{v'_{k'}}
   \draw (-4,-3) node[tre] (i_1) {}; \etq{i_1}
    \draw (-4,-5) node[tre] (i_2) {}; \etq{i_2}
    \draw (-4,-7.5) node[tre] (i_k) {}; \etq{i_k}
 \draw (4,-3) node[tre] (i'_1) {}; \etq{i'_1}
    \draw (4,-5) node[tre] (i'_2) {}; \etq{i'_2}
    \draw (4,-7.5) node[tre] (i'_{k'}) {}; \etq{i'_{k'}}
    \draw[->](xx)--(u);
        \draw[->](u) to (v);
    \draw[->](u) to (v_1);
        \draw[->](u) to (v'_1);
    \draw[->](v_1) to (i_1);
    \draw[->](v_1) to (v_2);
    \draw[->](v_2) to (i_2);
    \draw[->](v_k) to (i_k);
    \draw[->](v_k) to (i);    
    \draw[->](v'_1) to (i'_1);
    \draw[->](v'_1) to (v'_2);
     \draw[->](v'_2) to (i'_2);
   \draw[->](v'_{k'}) to (i'_{k'});
    \draw[->](v'_{k'}) to (i);    
  \end{tikzpicture}
               \begin{tikzpicture}[thick,>=stealth,scale=0.3]
              \draw(0,2) node{\ };
              \draw(0,0) node  {$\Longrightarrow$};
              \draw(0,-8.5) node{\ };
 \end{tikzpicture}
  \begin{tikzpicture}[thick,>=stealth,scale=0.3]
     \draw (0,2) node[tri] (xx) {};
    \draw (0,0) node[tri] (u) {}; \etq{u}
    \draw (-2,-2) node[tre] (i_1) {}; \etq{i_1}
       \draw (0,-2) node[tre] (i) {}; \etq{i}
       \draw (2,-2) node[tri] (v) {};
       \draw (3.5,-2) node  {$\ldots$};
  \draw (0,-8.5) node  {\ };
   \draw[->](xx) to (u);
      \draw[->](u) to (i);
      \draw[->](u) to (i_1);
      \draw[->](u) to (v);

     \end{tikzpicture}

\caption{\label{fig:G-red-1} 
The $\ovG_{i;i_1,\dots,i_k;i'_1,\dots,i'_{k'}}$ reduction.}
\end{figure}

\begin{remark}
In the $G$ and $\ovG$ reductions, we leave two tree leaves attached to
the former split node of the removed reticulation cycle in order to
ensure that their application never generates an out-degree 1 tree
node, while avoiding to increase unnecessarily the number of
reductions.
\end{remark}

Now we have the following basic applicability results.

\begin{proposition}\label{thm:red}
Let $N$ be a 1-nested network with more than one leaf.  Then, at least
one $R$, $T$, $G$, or $\ovG$ reduction can be applied to $N$, and
the result is a 1-nested network.
\end{proposition}

\begin{proof}
If $N$ contains some internal node $v$ with at least two children that
are tree leaves, say $i$ and $j$, then we can apply to $N$ the
$R_{i;j}$ reduction, if the out-degree of $v$ is 2, or the $T_{i;j}$
reduction, if its out-degree is greater than 2.

Assume now that $N$ does not contain any internal node with more than
one tree leaf child: in particular, it does not contain any internal
tree node with all its children tree leaves.  Then, by Proposition
\ref{prop:basic-2h1n}, it contains a hybrid node $h$ with all its
children tree leaves (and therefore, by the current assumption on $N$,
$h$ either is a leaf itself or has out-degree 1), and such that all
the intermediate nodes in its reticulation cycle $K$ have all their
children outside $K$ tree leaves (and therefore each one of them has
exactly one child outside $K$, by Lemma \ref{1n no ih} and the current
assumption on $N$): let $i_1,\ldots, i_k$ and $i'_1, \ldots, i'_{k'}$,
with $k\geq k'$, be the tree leaf children of the intermediate nodes
of the two respective merge paths of $K$, listed in descending order
of their parents along the path.  Then, if $h$ has out-degree 1 and
its child is the tree leaf $i$, we can apply to $N$ the
$G_{i;i_1,\ldots,i_k;i_1',\ldots,i_{k'}'}$ reduction, while if $h$ is
the leaf $i$, we can apply to $N$ the
$\ovG_{i;i_1,\ldots,i_k;i_1',\ldots,i_{k'}'}$ reduction.

The fact that the result of the application of a $R$, $T$, $G$, or
$\ovG$ reduction to $N$ is again a 1-nested network has been discussed
in the definition of the reductions.
\end{proof}

\begin{corollary}\label{thm:red-sbin}
Let $N$ be a semibinary 1-nested network with more than one leaf.
Then, at least one $R$, $T$, or $G$ reduction can be applied to $N$,
and the result is a semibinary 1-nested network.
\end{corollary}

\begin{proof}
Since $N$ does not contain hybrid leaves, we cannot apply to it any
$\ovG$ reduction, and therefore, by Proposition \ref{thm:red}, we can
apply to it at least one $R$, $T$, or $G$ reduction.

Now, if we can apply a $R_{i;j}$ or $T_{i;j}$ reduction to $N$, the
common parent of the tree leaves $i$ and $j$ is a tree node, and if we
can apply a $G_{i;i_1,\ldots,i_k;i'_1,\ldots,i'_{k'}}$ reduction, the
split node of the reticulation cycle for the hybrid parent of $i$ is a
tree node (in both cases because hybrid nodes in $N$ have out-degree
1), and therefore neither application produces a hybrid node of
out-degree different from 1.
\end{proof}

\begin{corollary}\label{thm:red-bin}
Let $N$ be a binary 1-nested network with more than one leaf.  Then,
at least one $R$ or $G$ reduction can be applied to $N$, and the
result is a binary 1-nested network.
\end{corollary}

\begin{proof}
Since $N$ does not contain nodes with out-degree greater than 2, we
cannot apply to it any $T$ reduction, and thus, by Corollary
\ref{thm:red-sbin}, we can apply to it some $R$ or $G$ reduction.

Now, if we apply a $R$ reduction to $N$, we replace an internal tree
node with two tree leaf children by a tree leaf, and the result is
again binary.  And if we apply to $N$ a $G$ reduction, the split node
of the reticulation cycle we remove is, as in the semibinary case, a
tree node, and in this case moreover without any child outside the
reticulation cycle (because its out-degree must be 2), and after the
application of the reduction it is still a tree node of out-degree 2.
\end{proof}

We shall call the inverses of the $R$, $T$, $G$, and $\ovG$
reductions, respectively, the $R^{-1}$, $T^{-1}$, $G^{-1}$, and
$\ovG^{-1}$ \emph{expansions}, and we shall denote them by
$R_{i;j}^{-1}$, $T^{-1}_{i;j}$,
$G_{i;i_1,\ldots,i_k;i_1',\ldots,i_{k'}'}^{-1}$, and
$\ovG_{i;i_1,\ldots,i_k;i_1',\ldots,i_{k'}'}^{-1}$.  More
specifically, for every 1-nested network $N$:
\begin{itemize}

\item if $N$ contains a leaf labeled with $i$ but no leaf labeled with
$j$, then the $R_{i;j}^{-1}$ expansion can be applied to $N$, and
$R_{i;j}^{-1}(N)$ is obtained by unlabeling the leaf $i$ and adding to
it two tree leaf children labeled $i$ and $j$;

\item if $N$ contains a tree leaf labeled with $i$ that has some
sibling, but no leaf labeled with $j$, then the $T_{i;j}^{-1}$
expansion can be applied to $N$, and $T_{i;j}^{-1}(N)$ is obtained by
adding to the parent of the leaf $i$ a new tree leaf child labeled
with $j$;

\item if $N$ contains an internal node $u$ with two tree leaf children
$i_1,i$, but no leaf labeled with $i_2,\ldots,i_k,i_1',\ldots,i'_{k'}$
(with $k\geq k'$), then the
$G_{i;i_1,\ldots,i_k;i_1',\ldots,i_{k'}'}^{-1}$ expansion can be
applied to $N$, and $G_{i;i_1,\ldots,i_k;i_1',\ldots,i_{k'}'}^{-1}(N)$
is obtained by removing the leaves $i,i_1$ and their incoming arcs,
and then starting in $u$ two new internally disjoint paths with $k$
and $k'$, respectively, intermediate nodes and ending in the same
hybrid node $h$, and then adding to each intermediate node of these
paths one new tree leaf and labeling these leaves (in descending order
along the paths) with $i_1,\ldots,i_k$ and $i_1',\ldots,i'_{k'}$,
respectively, and finally adding to $h$ a new tree leaf child labeled
with $i$;

\item the application condition for the
$\ovG_{i;i_1,\ldots,i_k;i_1',\ldots,i_{k'}'}^{-1}$ expansion is
exactly the same as for
$G_{i;i_1,\ldots,i_k;i_1',\ldots,i_{k'}'}^{-1}$, and
$\ovG_{i;i_1,\ldots,i_k;i_1',\ldots,i_{k'}'}^{-1}(N)$ is as
$G_{i;i_1,\ldots,i_k;i_1',\ldots,i_{k'}'}^{-1}(N)$, except that the
new hybrid node is itself a leaf labeled with $i$.
\end{itemize}
>From these descriptions we easily see that the result of a $R^{-1}$,
$T^{-1}$, $G^{-1}$ or $\ovG^{-1}$ expansion applied to a 1-nested
network is always a 1-nested network.

The following result is easily deduced from the explicit descriptions
of the reductions and expansions.

\begin{lemma}
\label{lem:iso-red}
Let $N$ and $N'$ be two 1-nested networks.  If $N\cong N'$, then the
result of applying to both $N$ and $N'$ the same $R^{-1}$ expansion
(respectively, $T^{-1}$ expansion, $G^{-1}$ expansion or $\ovG^{-1}$
expansion) are again two isomorphic 1-nested networks.

Moreover, if we apply a $R$, $T$, $G$, or $\ovG$ reduction to a
1-nested network $N$, then we can apply to the resulting network the
corresponding inverse $R^{-1}$, $T^{-1}$, $G^{-1}$, or $\ovG^{-1}$
expansion and the result is a 1-nested network isomorphic to $N$.
\qed
\end{lemma}

\section{Proving metrics through reductions}
\label{sec:proving}

Let $\mathcal{C}$ be throughout this section a class endowed with a
notion of isomorphism $\cong$.  A \emph{metric} on $\mathcal{C}$ is a
mapping
$$
d:\mathcal{C}\times \mathcal{C} \to \RR
$$
satisfying the following axioms: for every $A,B,C\in \mathcal{C}$,
\begin{enumerate}[(a)]

\item \emph{Non-negativity}: $d(A,B)\geq 0$;

\item \emph{Separation}: $d(A,B)=0$ if and only if $A\cong B$;

\item \emph{Symmetry}: $d(A,B)=d(B,A)$;

\item \emph{Triangle inequality}: $d(A,C)\leq d(A,B)+d(B,C)$.
\end{enumerate}
A \emph{metric space} is a pair $(X,d)$ where $X$ is a set and $d$ is
a metric on $X$, taking as the notion of isomorphism in $X$ the
equality (that is, replacing $\cong$ by $=$ in the separation axiom).

All distances for hybridization networks considered in this paper are
\emph{induced through representations}, in the following sense.  A
\emph{representation} of $\mathcal{C}$ in a metric space $(X,d)$ is a
mapping
$$
F:\mathcal{C}\to X
$$
such that if $A\cong B$, then $F(A)=F(B)$.

Given such a representation, the \emph{distance induced} by $d$
through $F$ is the mapping
$$
d_F:\mathcal{C}\times\mathcal{C}\to \RR
$$
defined by $d_F(A,B)=d(F(A),F(B))$, for every $A,B\in \mathcal{C}$.

The metric axioms for $d$ imply that this mapping is non-negative,
symmetric, it sends pairs of isomorphic members of $\cC$ to 0, and it
satisfies the triangle inequality.  So, to be a metric on $\cC$, $d_F$
only needs to satisfy that $d_F(A,B)=0$ implies $A\cong B$.  Now, it
is straightforward to prove the following result (cf.
\cite[Prop.~1]{cardona.ea:09-nak}).

\begin{lemma}\label{prop:metrics}
The mapping $d_F$ is a metric on $\mathcal{C}$ if, and only if, it is
injective up to isomorphism, in the sense that, for every $A,B\in
\mathcal{C}$, if $F(A)=F(B)$, then $A\cong B$.  \qed
\end{lemma}

Reductions as those introduced in the last section can be used to
prove the injectivity up to isomorphism of a representation $F$ and
hence, as a consequence, that the corresponding $d_F$ is a metric; it
was done for specific classes $\mathcal{C}$ of evolutionary networks
and specific metrics in
\cite{cardona.ea:sbtstc:2008,cardona.ea:tcbb:comparison.2:2008}.
Since we shall use several times this kind of proofs in this paper, we
make explicit here their general outline and the lemma they rely on.

Let $\mathcal{C}_{S',m}$ denote a class of 1-nested hybridization
networks of some specific type on a given set $S'$ and with at most
$m$ nodes, and let $\mathcal{C}_{S'}=\bigcup_{m\geq |S'|}
\mathcal{C}_{S',m}$.  Assume we have a set of reductions
$R_1,\ldots,R_s$ that can be applied to members of $\mathcal{C}_{S'}$,
with inverse expansions $R_1^{-1},\ldots, R_s^{-1}$.  Consider the
following conditions on these reductions and expansions:
\begin{itemize}

\item[(R1)] For every $N\in \mathcal{C}_{S',m}$ with $|S'|\geq 2$,
there exists some reduction $R_i$ that can be applied to $N$.

\item[(R2)] For every $N\in \mathcal{C}_{S',m}$ and for every
reduction $R_i$, $R_i(N)\in \mathcal{C}_{S'_i,m_i}$ for some
$S'_i\subseteq S'$ and $m_i< m$; moreover, $S'_i$ and $m_i$ only
depend on $S'$, $m$ and $R_i$, not on $N$.

\item[(R3)] For every $N\in \mathcal{C}_{S',m}$ and for every
reduction $R_i$, if $R_i$ can be applied to $N$, then $R_i^{-1}$ can
be applied to $R_i(N)$ and $R_i^{-1}(R_i(N))\cong N$.

\item[(R4)] For every reduction $R_i$ and for every $N,N'\in
\mathcal{C}_{S'_i,m_i}$ such that $N\cong N'$, if the corresponding
expansion $R_i^{-1}$ can be applied to $N$, then it can also be
applied to $N'$ and the resulting networks are isomorphic.
\end{itemize}

The definitions and results given in Section \ref{sec:red-1n} imply
that:
\begin{itemize}

\item The set of all $R$ and $G$ reductions satisfy conditions (R1) to
(R4) for the classes $\cC_{S'}$ of all binary 1-nested hybridization
networks on a set $S'$.

\item The set of all $R$, $T$, and $G$ reductions satisfy conditions
(R1) to (R4) for the classes $\cC_{S'}$ of all semibinary 1-nested
hybridization networks on a set $S'$.

\item The set of all $R$, $T$, $G$, and $\ovG$ reductions satisfy
conditions (R1) to (R4) for the classes $\cC_{S'}$ of all 1-nested
hybridization networks on a set $S'$.
\end{itemize}

Now, we have the following result.

\begin{lemma}\label{lem:basic-met}
Let $S$ be a given set of labels.  For every $S'\subseteq S$, let
$\mathcal{C}_{S',m}$ and $R_1,\ldots,R_s$ be as above, and assume that
these reductions satisfy conditions (R1) to (R4).  For every
$S'\subseteq S$, let $F_{S'}:\mathcal{C}_{S'}\to X_{S'}$ be a
representation in a metric space $(X_{S'},d^{(S')})$.

Then, $F_S$ is injective up to isomorphism if the following two
conditions are satisfied for every $S'\subseteq S$, for every $m\geq
|S'|$, for every reduction $R_i$, and for every $N,N'\in
\cC_{S',m}$ such that $F_{S'}(N)=F_{S'}(N')$:
\begin{itemize}

\item[(A)] If $R_i$ can be applied to $N$, then it can also be applied
to $N'$.

\item[(R)] If $R_i$ is applied to $N$ and $N'$, then
$F_{S'_i}(R_i(N))=F_{S'_i}(R_i(N'))$.
\end{itemize}

In particular, if these two conditions are satisfied, then
$d^{(S)}_{F_S}$ is a metric on $\mathcal{C}_S$.
\end{lemma}

\begin{proof}
We shall prove by induction on $|S'|+m$ the following statement:
\begin{quote}
For every $S'\subseteq S$ and $m\geq |S'|$, if $N,N'\in
\mathcal{C}_{S',m}$ are such that $F_{S'}(N)= F_{S'}(N')$, then
$N\cong N'$.
\end{quote}

The starting case, when $|S'|+m=2$, is obvious because then $S'$ must
be a singleton, and there is, up to isomorphism, only one 1-nested
hybridization network on a given singleton $\{i\}$: a single node
labeled with $i$.

Let now $N,N'\in \mathcal{C}_{S',m}$ be such that
$F_{S'}(N)=F_{S'}(N')$ and $|S'|+m\geq 3$.  If $|S'|=1$, we reason as
in the starting case to deduce that $N\cong N'$, so we assume that
$|S'|\geq 2$.  By (R1), some reduction $R_i$ can be applied to $N$,
and since $F_{S'}(N)=F_{S'}(N')$, by (A) it can also be applied to
$N'$.  Then, by (R2), $R_i(N), R_i(N')\in \mathcal{C}_{S'_i,m_i}$,
with $S'_i\subseteq S'$ and $m_i<m$, and by (R) we have that
$F_{S'_i}(R_i(N))=F_{S'_i}(R_i(N'))$.  Therefore the induction
hypothesis applies, implying that $R_i(N)\cong R_i(N')$.  But then, by
(R3), $R_i^{-1}$ can be applied to $R_i(N)$ and $R_i(N')$ and
$R_i^{-1}(R_i(N))\cong N$ and $R_i^{-1}(R_i(N'))\cong N'$, while, by
(R4), $R_i^{-1}(R_i(N))\cong R_i^{-1}(R_i(N'))$.  This implies that
$N\cong N'$, as we wanted to prove.

Thus, in particular, we have that for every $m\geq |S|$, if $N,N'\in
\mathcal{C}_{S,m}$ are such that $F_{S}(N)= F_{S}(N')$, then $N\cong
N'$.  Now notice that if $N,N'\in \mathcal{C}_S$, then there exists
some $m$ such that $N,N'\in \mathcal{C}_{S,m}$: take as $m$ the
largest number of nodes in $N$ or in $N'$.  Therefore, $F_S$ is
injective up to isomorphism, as we claimed.
\end{proof}

\begin{remark}
If one wants to use a result like the last lemma to prove the
injectivity up to isomorphism of a certain representation of $S$-rDAGs
more general than 1-nested networks, then it may be necessary to
explicitly add to (A) and (R) a third condition that covers the
starting case:
\begin{itemize}
\item[(S)] For every $i\in S$, $F_{\{i\}}$ is injective up to
isomorphism.
\end{itemize}
\end{remark}

We shall also use a couple of times the following straightforward
fact.

\begin{lemma}\label{lem:refines}
Let $F:\cC\to X$ and $F':\cC\to X'$ be two representations of $\cC$ in
metric spaces $(X,d)$ and $(X',d')$, and assume that $F(A)=F(B)$
implies $F'(A)=F'(B)$.  Then, if $F'$ is injective up to isomorphism,
so is $F$.  \qed
\end{lemma}

When the hypothesis of this lemma is satisfied, we say that $F$
\emph{refines} $F'$, and also that $d_F$ \emph{refines} $d'_{F'}$.
Notice that if $d'_{F'}$ is a metric and $d_F$ refines it, then it is
also a metric.

\section{Robinson-Foulds distance}
\label{sec:RF}

Let $N=(V,E)$ be a $S$-rDAG. For every node $v\in V$, the
\emph{cluster} of $v$ in $N$ is the set $C(v)\subseteq S$ of leaves
that are descendants of $v$.  The \emph{cluster representation} of $N$
is the multiset
$$
\cC(N)=\{C_N(v)\mid v\in V\},
$$
where each member appears with multiplicity the number of nodes having
it as cluster.  In particular, the cardinal of $\cC(N)$ (as a
multiset, that is, every element counted with its multiplicity) is
equal to the number of nodes in $N$.

The \emph{Robinson-Foulds distance} between a pair of $S$-rDAGs $N,N'$
is
$$
d_{RF}(N,N')=|\cC(N)\bigtriangleup \cC(N')|,
$$
where the symmetric difference and its cardinal refer to multisets.
It is the natural generalization to $S$-rDAGs of the well known
Robinson-Foulds distance for phylogenetic trees
\cite{robinson.foulds:1981}.

\begin{remark}\label{rem:gtnoreg}
If $v$ is an ancestor of $u$ in $N$, then $C(u)\subseteq C(v)$, but
the converse implication is false, even in binary galled trees.  See,
for instance, Fig.~\ref{fig:rf3} below: in both networks, the root and
its tree child have the same cluster, but the root is not a descendant
of its child.
\end{remark}

It is known that the Robinson-Foulds distance is a metric on the class
of all \emph{regular} evolutionary networks on a given set $S$ (the
networks $N$ such that the mapping $v\mapsto C(v)$ induces an
isomorphism of directed graphs between $N$ and the Hasse diagram of
$(\mathcal{C}(N),\supseteq)$) \cite{baroni.ea:ac04} and on the class
of all tree-child phylogenetic networks on a given set $S$ that do not
contain any hybrid node with two parents connected by a path
\cite{cardona.ea:07a}.  Unfortunately, 1-nested networks, or even
binary galled trees, need not be regular (by Remark \ref{rem:gtnoreg})
and they can contain reticulation cycles where one merge path is a
single arc.  So, we cannot use those results to prove that the
Robinson-Foulds distance is a metric, even on the class of all
binary galled trees.
 
\begin{figure}[htb]
\begin{center}
\begin{tikzpicture}[thick,>=stealth,scale=0.3]
\draw (0,0) node[hyb] (1) {}; \etq 1
\draw (4,0) node[tre] (2) {}; \etq 2
\draw (2,2) node[tre] (a) {}; 
\draw (2,4)  node[tre] (r) {};  
\draw[->](r)--(a);
\draw [->](r)--(1);
\draw [->](a)--(1);
\draw [->](a)--(2);
\end{tikzpicture}
\qquad
\begin{tikzpicture}[thick,>=stealth,scale=0.3]
\draw (0,0) node[tre] (1) {}; \etq 1
\draw (4,0) node[hyb] (2) {}; \etq 2
\draw (2,2) node[tre] (a) {}; 
\draw (2,4)  node[tre] (r) {};  
\draw[->](r)--(a);
\draw [->](r)--(2);
\draw [->](a)--(1);
\draw [->](a)--(2);
\end{tikzpicture}
\end{center}
\caption{\label{fig:rf4} Two non-isomorphic galled trees with the same cluster representation and hybrid nodes of out-degree 0.}
\end{figure}
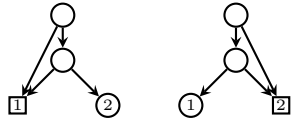

As a matter of fact, the cluster representation is not injective up to
isomorphism, and hence the Robinson-Foulds distance is not a metric,
for 1-nested networks, or even galled trees, unless we restrict the
possible in- and out-degrees of their nodes: they cannot contain
either internal tree nodes of out-degree other than 2 (see
Fig.~\ref{fig:rf1}), or hybrid nodes of out-degree 0 (see
Fig.~\ref{fig:rf4}) or greater than 1 (see Fig.~\ref{fig:rf3}).
Therefore, the Robinson-Foulds distance can only be a metric for
\emph{binary} 1-nested networks, that is, for binary galled trees.
Now, we have the following result.

\begin{figure}[htb]
\begin{center}
\begin{tikzpicture}[thick,>=stealth,scale=0.25]
\draw (0,0) node[tre] (1) {}; \etq 1
\draw (2,0) node[tre] (2) {}; \etq 2
\draw (4,0) node[tre] (3) {}; \etq 3
\draw (6,0) node[tre] (4) {}; \etq 4
\draw (4,2) node[hyb] (A) {};  
\draw (2,4) node[tre] (b) {};  
\draw (2,7) node[tre] (a) {};  
\draw (4,8) node[tre] (r) {};  
\draw[->](r)--(a);
\draw [->](r)--(4);
\draw [->](r)--(A);
\draw [->](a)--(1);
\draw [->](a)--(b);
\draw [->](b)--(2);
\draw [->](b)--(A);
\draw [->](A)--(3);
\end{tikzpicture}
\qquad
\begin{tikzpicture}[thick,>=stealth,scale=0.25]
\draw (0,0) node[tre] (1) {}; \etq 1
\draw (2,0) node[tre] (2) {}; \etq 2
\draw (4,0) node[tre] (3) {}; \etq 3
\draw (6,0) node[tre] (4) {}; \etq 4
\draw (4,2) node[hyb] (A) {};  
\draw (2,4) node[tre] (b) {};  
\draw (2,7) node[tre] (a) {};  
\draw (4,8) node[tre] (r) {};  
\draw[->](r)--(a);
\draw [->](r)--(4);
\draw [->](a)--(A);
\draw [->](a)--(1);
\draw [->](a)--(b);
\draw [->](b)--(2);
\draw [->](b)--(A);
\draw [->](A)--(3);
\end{tikzpicture}
\end{center}
\caption{\label{fig:rf1} Two non-isomorphic galled trees with the same cluster representation and internal tree nodes of out-degree 3.}
\end{figure}
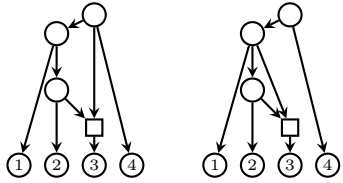

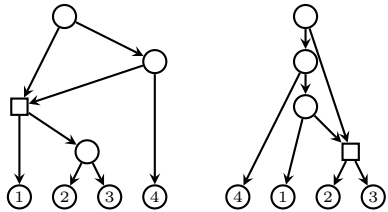
\begin{figure}[htb]
\begin{center}
\begin{tikzpicture}[thick,>=stealth,scale=0.3]
\draw (0,0) node[tre] (1) {}; \etq 1
\draw (2,0) node[tre] (2) {}; \etq 2
\draw (4,0) node[tre] (3) {}; \etq 3
\draw (6,0) node[tre] (4) {}; \etq 4
\draw (3,2) node[tre] (b) {}; 
\draw (0,4) node[hyb] (A) {}; 
\draw (6,6) node[tre] (a) {}; 
\draw (2,8) node[tre] (r) {}; 
\draw[->](r)--(a);
\draw [->](r)--(A);
\draw [->](a)--(4);
\draw [->](a)--(A);
\draw [->](A)--(1);
\draw [->](A)--(b);
\draw [->](b)--(2);
\draw [->](b)--(3);
\end{tikzpicture}
\qquad
\begin{tikzpicture}[thick,>=stealth,scale=0.3]
\draw (0,0) node[tre] (4) {}; \etq 4
\draw (2,0) node[tre] (1) {}; \etq 1
\draw (4,0) node[tre] (2) {}; \etq 2
\draw (6,0) node[tre] (3) {}; \etq 3
\draw (5,2) node[hyb] (A) {}; 
\draw (3,4) node[tre] (b) {}; 
\draw (3,6) node[tre] (a) {}; 
\draw (3,8) node[tre] (r) {}; 
\draw[->](r)--(a);
\draw [->](r)--(A);
\draw [->](a)--(4);
\draw [->](a)--(b);
\draw [->](b)--(1);
\draw [->](b)--(A);
\draw [->](A)--(2);
\draw [->](A)--(3);
\end{tikzpicture}
\end{center}
\caption{\label{fig:rf3} Two non-isomorphic galled trees with the same cluster representation and hybrid nodes of out-degree 2.}
\end{figure}

\begin{theorem}
\label{prop:RF-AR}
Let $N,N'$ be two binary 1-nested networks on a given set $S$ such
that $\cC(N)=\cC(N')$.
\begin{itemize}

\item[(A)] If a specific $R$ or $G$ reduction can be applied to $N$,
then it can also be applied to $N'$.

\item[(R)] If a specific $R$ or $G$ reduction is applied to $N$ and
$N'$, the resulting networks have the same cluster representations.
\qed
\end{itemize}
\end{theorem}

In order not to lose the thread of the paper, we postpone the proof of
this theorem until \S A2 in the Appendix at the end of the paper.
Combining Lemma \ref{lem:basic-met} with this theorem, we obtain the
following result.

\begin{corollary}\label{th:RF}
The Robinson-Foulds distance is a metric on the class of all binary
galled trees on a given set $S$.  \qed
\end{corollary}

\section{Tripartitions distance}
\label{sec:trip}

Let $N=(V,E)$ be a $S$-rDAG. For every node $v\in V$, let
$A(v)\subseteq S$ be the set of (labels of) strict descendant leaves
of $v$ and $B(v)= C(v)\setminus A(v)$ the set of non-strict descendant
leaves of $v$; $B(v)$ may be empty, but $A(v)\neq\emptyset$ by Lemma
\ref{1n no ih}.  The \emph{tripartition} associated to $v$
\cite{moret.ea:2004} is
$$
\theta(v)=(A(v),B(v),S\setminus C(v)).
$$
Notice that the tripartition associated to a node $v$ refines its
cluster $C(v)$, by splitting it into $A(v)$ and $B(v)$.

The \emph{tripartitions representation} of $N$ is the multiset
$$
\theta(N)=\{\theta(v)\mid v\in V\}
$$
of tripartitions of the nodes of $N$.  The \emph{tripartitions
distance} between a pair of $S$-rDAGs $N,N'$ is
$$
d_{tri}(N,N')=|\theta(N)\bigtriangleup \theta(N')|,
$$
where the symmetric difference and its cardinal refer to multisets.

It turns out that the tripartitions distance is a metric on the class
of all 1-nested networks on a given set.  It is a consequence of the
following proposition, whose proof we postpone until \S A3 in the
Appendix.

\begin{theorem}
\label{prop:tri-AR}
Let $N,N'$ be two 1-nested networks on a given set $S$ such that
$\theta(N)= \theta(N')$.
\begin{itemize}

\item[(A)] If a specific $R$, $T$, $G$, or $\ovG$ reduction can be
applied to $N$, then it can also be applied to $N'$.

\item[(R)] If a specific $R$, $T$, $G$, or $\ovG$ reduction is applied
to $N$ and $N'$, the resulting networks have the same tripartitions
representations.  \qed
\end{itemize}
\end{theorem}

So, using Lemma \ref{lem:basic-met}, we deduce the following result.

\begin{corollary}\label{th:tri}
The tripartitions distance is a metric on the class of all 1-nested
networks on a given set $S$.  \qed
\end{corollary}

\begin{remark}\label{rem:mu}
Another refinement (in the sense of Lemma \ref{lem:refines}) of the
Robinson-Foulds distance, the so-called \emph{$\mu$-distance}, was
introduced by Cardona \textsl{et al} \cite{cardona.ea:07b} and proved
to be a metric on the class of all tree-child $S$-rDAGs for any given
$S$: then, in particular, it is a metric on the class of all 1-nested
networks on a set $S$.  Soon later, L. Nakhleh \cite{nakhleh:2009}
proposed a distance $m$ that turned out to refine the
$\mu$-distance \cite{cardona.ea:09-nak} and therefore that is also a
metric on the class of all 1-nested networks on a set $S$.  The
interested reader can look up the aforementioned references for the
specific definitions of these metrics.
\end{remark}

\section{Nodal and splitted nodal distances}
\label{subsec:nodal}

Let $N=(V,E)$ be a $S$-rDAG; to simplify the language, throughout this
section we assume that $S=\{1,\ldots,n\}$ with $n=|S|$.  Recall from
\cite{cardona.ea:tcbb:comparison.1:2008} that the \emph{least common
semistrict ancestor}, LCSA for short, of a pair of nodes $u,v\in V$ is
the node that is a common ancestor of $u$ and $v$ and strict ancestor
of at least one of them, and that is a descendant of all other nodes
in $N$ satisfying these properties.  Such a LCSA of a pair of nodes
$u,v$ always exists and it is unique \cite[\S
IV]{cardona.ea:tcbb:comparison.1:2008}, and we shall denote it by
$[u,v]$.

The LCSA of a pair of nodes in a phylogenetic tree is their lowest
common ancestor.  It turns out that such a characterization extends to
1-nested networks.  Recall that a \emph{lowest common ancestor}, LCA
for short, of a pair of nodes $u,v$ in a rDAG is any common ancestor
of $u$ and $v$ that is not a proper ancestor of any other common
ancestor of them \cite{bender.ea:2005}.

\begin{lemma}\label{lem:lcsa-1n}
Every pair of nodes $u,v$ in a 1-nested network has only one LCA, and
it is their LCSA.
\end{lemma}

\begin{proof}
Let $x$ be any LCA of $u$ and $v$, and let us prove that $x$ must be a
strict ancestor of $u$ or $v$.  Indeed, by Lemma \ref{lem:RF-previ} in
\S A1 in the Appendix, if $x$ is not a strict ancestor of $u$, then it
is intermediate in the reticulation cycle for a hybrid node $h_u$ that
is a strict ancestor of $u$.  In a similar way, if $x$ is not a strict
ancestor of $v$, then it is intermediate in the reticulation cycle for
a hybrid node $h_v$ that is a strict ancestor of $v$.  Now, if $x$
were not a strict ancestor either of $u$ or of $v$, then it either
would happen that it is intermediate in reticulation cycles for two
different hybrid nodes, which is impossible in a 1-nested network, or
that it is a proper ancestor of a common ancestor of $u$ and $v$,
namely $h_u=h_v$, against the assumption that $x$ is a LCA of $u$ and
$v$.

So, $x$ is a common ancestor of $u$ and $v$ and a strict ancestor of
at least one of them, and thus it is an ancestor of $[u,v]$.  Since
$x$ cannot have proper descendants that are common ancestors of $u$
and $v$, we conclude that $x=[u,v]$.
\end{proof}

For every pair of leaves $i,j\in S$, let $\ell_N(i,j)$ and
$\ell_N(j,i)$ be the distances from $[i,j]$ to $i$ and to $j$,
respectively, and let $\nu_N(i,j)=\ell_N(i,j) +\ell_N(j,i)$.

The \emph{LCSA-path lengths matrix} of $N$ is the symmetric matrix
$$
\nu(N)=\left(\begin{array}{ccc} \nu_N(1,1) & \ldots & \nu_N(1,n)\\
\vdots & \ddots & \vdots\\
\nu_N(n,1) & \ldots & \nu_N(n,n)
\end{array}
\right)
$$
and the \emph{splitted LCSA-path lengths matrix} of $N$ is the (not
necessarily symmetric) matrix
$$
\ell(N)=\left(\begin{array}{ccc} \ell_N(1,1) & \ldots & \ell_N(1,n)\\
\vdots & \ddots & \vdots\\
\ell_N(n,1) & \ldots & \ell_N(n,n)
\end{array}
\right)
$$

The \emph{nodal distance} between a pair of $S$-rDAGs $N,N'$ is half
the Manhattan, or $L_1$, distance between $\nu(N)$ and $\nu(N')$:
$$
d_{\nu}(N,N')=\frac{1}{2} \sum_{1\leq i\neq j\leq n}
|\nu_N(i,j)-\nu_{N'}(i,j)|.
$$
The \emph{splitted nodal distance} between $N$ and $N'$ is the
Manhattan distance between $\ell(N)$ and $\ell(N')$:
$$
d_{\ell}(N,N')=\sum_{1\leq i\neq j\leq n}
|\ell_N(i,j)-\ell_{N'}(i,j)|.
$$
Of course, instead of using the Manhattan distance on the set of
$n\times n$ matrices, one can use any other distance for real-valued
matrices to compare LCSA-path lengths, or splitted LCSA-path lengths,
matrices, like for instance the euclidean distance.  The results in
this section do not depend on the actual metric for real-valued
matrices used.

The nodal distance $d_{\nu}$ is the natural generalization to
$S$-rDAGs of the classical nodal metric for binary phylogenetic trees
\cite{farris:sz73,willcliff:taxon71}, while the splitted nodal
distance $d_{\ell}$ generalizes to $S$-rDAGs the recently introduced
homonymous metric for arbitrary phylogenetic trees
\cite{cardona.ea:jmb:2008}.

It is known \cite{nodalnetworks,cardona.ea:tcbb:comparison.2:2008}
that $d_{\nu}$ is a metric on the class of all binary tree-child time
consistent phylogenetic networks on a given set $S$, and that $d_\ell$
is a metric on the class of all tree-child time consistent
phylogenetic networks on a given set $S$, but no binary galled tree
containing a reticulation cycle with one merge path consisting of a
single arc is time consistent, and therefore we cannot use these
results to prove that $d_{\nu}$ or $d_\ell$ are metrics even for
binary galled trees.

\begin{figure}[htb]
\begin{center}
\begin{tikzpicture}[thick,>=stealth,scale=0.3]
\draw (0,0) node[tre] (1) {}; \etq 1
\draw (3,2) node[hyb] (A) {};  
\draw (3,0) node[tre] (2) {}; \etq 2
\draw (1,3) node[tre] (a) {}; 
\draw (2,5)  node[tre] (r) {};  
\draw[->](r)--(a);
\draw [->](r)--(A);
\draw [->](a)--(1);
\draw [->](a)--(A);
\draw [->](A)--(2);
\end{tikzpicture}
\qquad
\begin{tikzpicture}[thick,>=stealth,scale=0.3]
\draw (0,0) node[tre] (2) {}; \etq 2
\draw (3,2) node[hyb] (A) {}; \etq 2
\draw (3,0) node[tre] (1) {}; \etq 1
\draw (1,3) node[tre] (a) {}; 
\draw (2,5)  node[tre] (r) {};  
\draw[->](r)--(a);
\draw [->](r)--(A);
\draw [->](a)--(2);
\draw [->](a)--(A);
\draw [->](A)--(1);
\end{tikzpicture}
\end{center}
\caption{\label{fig:nod0} Two non-isomorphic binary galled trees on $S=\{1,2\}$ and whith the same LCSA-path length, 3, between their only two leaves.}
\end{figure}
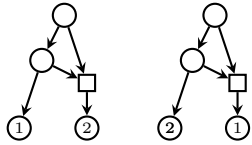

\begin{figure}[htb]
\begin{center}
\begin{tikzpicture}[thick,>=stealth,scale=0.3]
\draw (0,0) node[tre] (1) {}; \etq 1
\draw (4,0) node[tre] (2) {}; \etq 2
\draw (2,4)  node[tre] (r) {};  
\draw[->](r)--(1);
\draw [->](r)--(2);
\end{tikzpicture}
\qquad
\begin{tikzpicture}[thick,>=stealth,scale=0.3]
\draw (0,0) node[tre] (1) {}; \etq 1
\draw (4,0) node[hyb] (2) {}; \etq 2
\draw (1,2) node[tre] (a) {}; 
\draw (2,4)  node[tre] (r) {};  
\draw[->](r)--(a);
\draw [->](r)--(2);
\draw [->](a)--(1);
\draw [->](a)--(2);
\end{tikzpicture}
\end{center}
\caption{\label{fig:nod1} Two non-isomorphic galled trees with the same $\ell$ matrix and hybrid nodes of out-degree 0.}
\end{figure}
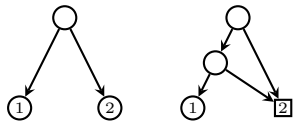

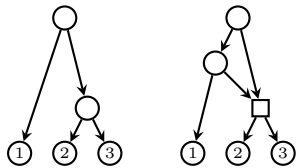
\begin{figure}[htb]
\begin{center}
\begin{tikzpicture}[thick,>=stealth,scale=0.3]
\draw (0,0) node[tre] (1) {}; \etq 1
\draw (2,0) node[tre] (2) {}; \etq 2
\draw (4,0) node[tre] (3) {}; \etq 3
\draw (3,2)  node[tre] (a) {};  
\draw (2,6)  node[tre] (r) {};  
\draw[->](r)--(1);
\draw [->](r)--(a);
\draw [->](a)--(2);
\draw [->](a)--(3);
\end{tikzpicture}
\qquad
\begin{tikzpicture}[thick,>=stealth,scale=0.3]
\draw (0,0) node[tre] (1) {}; \etq 1
\draw (2,0) node[tre] (2) {}; \etq 2
\draw (4,0) node[tre] (3) {}; \etq 3
\draw (3,2) node[hyb] (A) {};
\draw (1,4) node[tre] (a) {}; 
\draw (2,6)  node[tre] (r) {};  
\draw[->](r)--(a);
\draw [->](r)--(A);
\draw [->](a)--(1);
\draw [->](a)--(A);
\draw [->](A)--(2);
\draw [->](A)--(3);
\end{tikzpicture}
\end{center}
\caption{\label{fig:nod2} Two non-isomorphic galled trees with the same $\ell$ matrix and hybrid nodes of out-degree greater than 1.}
\end{figure}

It turns out that $\nu$ is not injective up to isomorphism, and hence
$d_{\nu}$ is not a metric, even for binary galled trees, as
Fig.~\ref{fig:nod0} shows.  As far as $\ell$ goes, it is not injective
up to isomorphism for 1-nested networks, or even galled trees, that
are not semibinary: if we allow hybrid nodes of out-degree 0 (see
Fig.~\ref{fig:nod1}) or greater than 1 (see Fig.~\ref{fig:nod2}),
there exist pairs of non-isomorphic galled trees with the same
splitted LCSA-path length matrices.  Therefore, $d_\ell$ can be a
metric at most on the class of all semibinary 1-nested networks.  Now,
we have the following result.

\begin{theorem}
\label{prop:snod-AR}
Let $N,N'$ be two semibinary 1-nested networks on a given set $S$ such
that $\ell(N)= \ell(N')$.
\begin{itemize}

\item[(A)] If a specific $R$, $T$, or $G$ reduction can be applied to
$N$, then it can also be applied to $N'$.

\item[(R)] If a specific $R$, $T$, $G$ reduction is applied to $N$ and
$N'$, the resulting networks have the same splitted LCSA-path lengths
matrices.  \qed
\end{itemize}
\end{theorem}

As we did previously, we postpone the proof of this theorem until \S
A4 in the Appendix at the end of the paper.  Combining Lemma
\ref{lem:basic-met} with this theorem, we obtain the following result.

\begin{corollary}\label{th:nod}
The splitted nodal distance is a metric on the class of all semibinary
1-nested networks on a given set $S$.  \qed
\end{corollary}

\section{Conclusion}

Several slightly different definitions of galled tree, capturing the
notion of a hybridization network with isolated reticulation cycles,
have been proposed so far in the literature.  The most general such
definition is as a network with arc-disjoint reticulation cycles
\cite{CJSS2005,jansson.sung:2006}, called in this paper
\emph{1-nested}, and the most restrictive is Gusfield \textsl{et al}'s
original definition of a galled tree as a network with node-disjoint
reticulation cycles \cite{gusfield.ea:csb:2003}: in between lie the
level-1 networks of Janson, Sung \textsl{et al}
\cite{JanssonSung2004a,JanssonSung2008}.  In the \emph{semibinary}
(hybrid nodes of in-degree 2 and out-degree 1) case, level-1 and
1-nested networks are the same, and in the \emph{binary} (semibinary
plus tree nodes of out-degree 2) case, galled trees, level-1 networks
and 1-nested networks are the same objects.

In this paper we have established for which classes of 1-nested
networks on a fixed set of labels, several distance measures
introduced so far in the literature satisfy the axioms of metrics:
actually, only the separation axiom (distance 0 means isomorphism) is
relevant here, because all other axioms of metrics are always
satisfied by these distances.  In summary, we have proved that:

\begin{enumerate}[(a)]
\item The Robinson-Foulds distance
\cite{baroni.ea:ac04,cardona.ea:tcbb:comparison.1:2008} is a metric
only for binary galled trees.

\item The tripartitions distance \cite{moret.ea:2004}, the
$\mu$-distance \cite{cardona.ea:07b} and Nakhleh's metric $m$ for
reduced networks \cite{nakhleh:2009} are metrics for arbitrary
1-nested networks.

\item The natural translation of the nodal distance for phylogenetic
trees to evolutionary networks \cite{nodalnetworks} is not a metric
even for binary galled trees.

\item The splitted nodal distance
\cite{nodalnetworks,cardona.ea:tcbb:comparison.2:2008} is a metric for
semibinary 1-nested networks, but not for
arbitrary galled trees.
\end{enumerate}
We would like to mention that the 1-nested networks turn out to
form the first well-defined class of evolutionary networks where
the tripartitions distance is shown to be a metric but the
Robinson-Foulds distance is not a metric.

There are other distances that have not been discussed in this paper
because they obviously fail to be metrics even for binary galled
trees.  This is the case of the triplets distance
\cite{cardona.ea:tcbb:comparison.2:2008}, which cannot be a metric for
binary galled trees because there are many more binary galled trees
with 3 leaves than possible triplets in the sense defined in the
aforementioned paper.  And, as it was already observed in \cite[\S
II.D]{cardona.ea:07b}, it is also the case of any distance defined by
comparing the multisets of induced subtrees, or the multisets of
splits of induced subtrees: for instance, the pairs of galled trees
depicted in Figs.  \ref{fig:nod1} or \ref{fig:nod2} have the same
multisets of induced subtrees.

The splitted nodal distance and the triplets distance were introduced
in \cite{cardona.ea:tcbb:comparison.2:2008} as suitable
generalizations of the corresponding distances for phylogenetic
networks with the aim of obtaining metrics on the class of tree-child
time consistent phylogenetic networks, and hence they were not
designed to cope with reticulation cycles where one merge path is a
single arc.  This is the main reason of their failure as metrics for
arbitrary 1-nested networks.  But it seems not difficult to modify
them to obtain metrics for 1-nested networks, by taking into
account the restricted, and specific, topological structure of these
networks: something similar was already done with the splitted nodal
distance to make it work on tree-child time consistent evolutionary
networks with hybrid nodes of (almost) arbitrary type
\cite{nodalnetworks}.

Galled trees, 1-nested networks, and level-1 networks are defined as having hybrid nodes of in-degree 2, in the first case by semantical reasons and in the other two cases for practical reasons (to guarantee that certain reconstruction algorithms run in polynomial time), and we have kept this restriction in this paper. But, although Gusfield \textsl{et al}'s node-disjoint reticulation cycles condition implies that hybrid nodes must have in-degree 2,  this restriction is not necessary in level-1 and 1-nested networks, and polynomial time algorithms for the reconstruction of level-1 or 1-nested networks with hybrid nodes of arbitrary in-degree may be discovered in the future, in which case it would be interesting to know whether the distance measures discussed in this paper define metrics in this more general case and they can be used thus to assess these new algorithms.

\section*{Appendix: Proofs of the main theorems}

\subsection*{A1\quad Some lemmas on clusters and tripartitions}

We establish in this subsection some basic properties of clusters on
1-nested networks that will be used in the proofs of the next two
subsections.  To simplify the notations, given a 1-nested network $N$
on a set $S$, let $\cIC(N)$ denote the multiset of clusters of its
internal nodes.  $\cIC(N)$ is obtained by removing from $\cC(N)$ one
copy of every singleton $\{i\}$ with $i\in S$.

\begin{lemma}
\label{lem:RF-previ}
Let $N$ be a 1-nested network on $S$.

\begin{enumerate}[(a)]
\item For every $i\in S$ and for every internal node $v$, $C(v)=\{i\}$
if, and only if, $i$ is a tree leaf and $v$ is its parent and it has
out-degree 1.

\item If two leaves $i,j$ are such that there does not exist any
member of $\cIC(N)$ containing one of them and not the other, then
they are sibling.

\item Let $v$ be a tree node and $u$ its only parent.  If $C(u)\neq
C(v)$, then $C(u)$ is the only (up to multiplicities) minimal
member of $\cIC(N)$ strictly containing $C(v)$.  If $C(u)=C(v)$ and
$u$ has out-degree greater than 1, then $u$ is the split node of a
reticulation cycle such that one of the merge paths contains $v$ as
intermediate node and the other merge path is a single arc.

\item If a node $v$ is a non-strict descendant of a node $u$, then $u$
is intermediate in the reticulation cycle for a hybrid node that is a
strict ancestor of $v$.
\end{enumerate}
\end{lemma}

\begin{proof}
\emph{(a)} If $v$ is a node with only one child and this child is the
tree leaf $i$, then $C(v)=\{i\}\in \cIC(N)$.  Conversely, let $v$ be
an internal node such that $C(v)=\{i\}$.  Since every internal node
in $N$ has a tree descendant leaf, $i$ must be a tree descendant
leaf of $v$ (and in particular a tree leaf).  Let $w$ be the parent of
$i$, and let us prove that it has out-degree 1.  Indeed, if $u$ is a
child of $w$ other than $i$, it has a tree descendant leaf $j$, and
$j\neq i$, because, otherwise, the only parent $w$ of $i$ would be a
descendant of its child $u$.  But then $j\in C(v)$, against the
assumption that $C(v)=\{i\}$.

So, the tree path $v\pathgr i$ cannot have any intermediate node,
because otherwise $w$ would be intermediate in this path and hence it
would be a tree node, but internal tree nodes in $N$ have out-degree greater
than 1.  Therefore, $v$ is the parent of $i$.  But then, as we have
just seen, it must have out-degree 1.  \smallskip

\emph{(b)} Assume that the every member of $\cIC(N)$ containing
$i$ or $j$ contains both of them, but that $i$ and $j$ are not
sibling.  Let $v_1$ be a parent of $i$: then $i\in C(v_1)$ implies
$j\in C(v_1)$ and, since $v_1$ is not a parent of $j$, $v_1$ is a
proper ancestor of some parent $w_1$ of $j$.  Then, $j\in C(w_1)$
implies $i\in C(w_1)$ and thus, since $w_1$ is not a parent of $i$,
$w_1$ is a proper ancestor of some parent $v_2\neq v_1$ (because $v_1$
is a proper ancestor of $w_1$) of $i$.  Iterating this process, we
obtain that $v_2$ is a proper ancestor of another parent $w_2\neq w_1$
of $j$, and then that $w_2$ is a proper ancestor of another parent
$v_3\neq v_1,v_2$ of $i$, which is impossible because every node in
$N$ has at most 2 parents. \smallskip

%


\emph{(c)} Let $u$ be the parent of the tree node $v$.  Assume that
$C(u)\neq C(v)$ and let $i$ be a tree descendant leaf of $v$, and
hence also of $u$.  For every other internal node $w$, if
$C(v)\subsetneq C(w)$, then $i\in C(w)$, and therefore either the path
$u\pathgr i$ is contained in the path $w\pathgr i$ or conversely.  But
$C(v)\subsetneq C(w)$ implies that $w$ cannot be a descendant of $v$,
and we conclude that $u\pathgr i$ is contained in the path
$w\pathgr i$, and hence $u$ is a descendant of $w$, which implies that
$C(u)\subseteq C(w)$.

Assume now that $u$ has out-degree greater than 1 and that
$C(u)=C(v)$.  Let $v'$ be another child of $u$ and let $j$ be a tree
descendant leaf of $v'$.  Then, $j\in C(v)$, and therefore either the
path $v\pathgr j$ contains the path $v'\pathgr j$ or the path
$v'\pathgr j$ contains the path $v\pathgr j$.  But the last situation
is impossible, because if $v$ belongs to the path $v'\pathgr j$, its
only parent $u$ should also belong to it, and $u$ cannot be a
descendant of its child $v'$.  So, we conclude that $v'$ is a
descendant of $v$, and therefore that $v'$ is a hybrid node and its
reticulation cycle consists of the arc $(u,v')$ and a path
$(u,v,\ldots,v')$.  \smallskip

\noindent\emph{(d)} Assume that $v$ is a non-strict descendant of $u$.
Let $u\pathgr v$ be any path from $u$ to $v$, and $r\pathgr v$ a path
from the root $r$ of $N$ to $v$ not containing $u$.  Let $w$ be the
first node in $u\pathgr v$ contained also in $r\pathgr v$.  Since, by
assumption, $w\neq u$ and, clearly, $w\neq r$, $w$ will have different
parents in both paths, which implies that it is hybrid.  Let now
$r\pathgr u$ be any path from the root to $u$ and let $x$ be the last
node in this path belonging to the subpath $r\pathgr w$ of $r\pathgr
v$: again, $u\neq x$.  Then, the subpath $x\pathgr w$ of $r\pathgr v$
and the concatenation of the subpath $x\pathgr u$ of $r\pathgr u$ and
the subpath $u\pathgr w$ of $u\pathgr v$ are internally disjoint, and
hence they form a reticulation cycle for $w$ with split node $x$ and
having $u$ as intermediate node.

It remains to prove that $w$ is a strict ancestor of $v$.  But if it
were not, then, as we have just seen, $w$ would be intermediate in a
reticulation cycle for an ancestor of $v$, which is impossible by
Lemma \ref{1n no ih}.
\end{proof}

\begin{lemma}
\label{lem:1n-cicles}
Let $N$ be a 1-nested network on $S$, let $h$ be a hybrid node of $N$
with $C(h)=\{i\}$, and let $K$ be its reticulation cycle, with split
node $u$.
\begin{enumerate}[(a)]

\item No pair of intermediate nodes of $K$ in different merge paths
are connected by a path.

\item Every pair of intermediate nodes in $K$ have different clusters,
and different also from $C(h)$.

\item The only non-strict descendant of each intermediate node of $K$
is $i$.

\item The intersection of the clusters of any pair of intermediate
nodes of different merge paths of $K$ is $\{i\}$.

\item $i$ is a strict descendant of $u$.

\item If $v$ is a node outside $K$ such that $i\in C(v)$, then $v$ is
an ancestor of $u$ and thus $C(u)\subseteq C(v)$.

\item All clusters of intermediate nodes in $K$ have multiplicity 1 in
$\cIC(N)$, except the cluster of the child other than $h$ of
$u$ when one of the merge paths consists of a single arc $(u,h)$.

\item The minimal elements of $\cIC(N)$ strictly containing $C(h)$ are
the clusters of the parents of $h$ that are intermediate in $K$.

\end{enumerate}
\end{lemma}

\begin{proof}
By Lemma \ref{lem:RF-previ}.(a), $C(h)=\{i\}$ implies that either
$h=i$ or that $i$ is a tree child of $h$, and its only child.
\smallskip

\noindent \emph{(a)} If $x$ and $y$ were two intermediate nodes
of $K$ belonging to different merge paths and there existed a path
$x\pathgr y$, then the first node in this path also belonging to the
path $u\pathgr y$ would have different parents in both paths, and
therefore it would be hybrid, which is impossible by Lemma~\ref{1n no ih}.
\smallskip

\noindent \emph{(b)} Let $x$ and $y$ be two different intermediate
nodes of $K$: if they belong to the same merge path, we take them so
that $y$ is a proper descendant of $x$.  We shall prove that $C(x)\neq
C(y)$.

Since both nodes are of tree type, $x$ has a child $v$ outside $K$.
Let $l$ be a tree descendant leaf of $v$, and assume that $l\in C(y)$.
Then, either the path $v\pathgr l$ contains the path $y\pathgr l$ or
vice versa.  But the tree path $u\pathgr y$ contained in the merge
path is the unique path from $u$ to $y$, and it does not contain $v$,
and therefore $y$ cannot be a descendant of $v$.  Thus, $v$ is a
descendant of $y$, and since $x$ is not a descendant of $y$ by (a), we
conclude that $v$ is a hybrid node such that its parent other than $x$
is a descendant of $y$.  But then, $y$ is intermediate in the
reticulation cycle of $v$, which is impossible because it is already
intermediate in the reticulation cycle of $h$.  So, we reach a
contradiction that implies that $l\notin C(y)$, and hence that
$C(x)\neq C(y)$.

On the other hand, Lemma \ref{lem:RF-previ}.(a) implies that, for
every proper ancestor $x$ of $h$, $C(h)=\{i\}\subsetneq C(x)$.
\smallskip

\noindent \emph{(c)} Let $x$ be an intermediate node of $K$ and $l$ a
descendant leaf of $x$ other than $i$.  If $l$ were a non-strict
descendant of $x$, then $x$ would be intermediate in the reticulation
cycle of a hybrid ancestor of $l$ by Lemma \ref{lem:RF-previ}.(d),
which is impossible because $x$ is already intermediate in $K$ and $h$
is not an ancestor of $l$.  Thus, every descendant leaf of $x$ other
than $i$ is a strict descendant of $x$.

On the other hand, the fact that $i$ is a non-strict descendant of $x$
is obvious: the composition of any path $r\pathgr u$ with the merge
path $u\pathgr h$ not containing $x$, and ending, if necessary, with
the arc $(h,i)$, yields a path $r\pathgr i$ not containing $x$.
\smallskip

\noindent \emph{(d)} Let $x$ and $y$ be two intermediate nodes of
different merge paths of $K$.  If there existed some leaf $l\neq i$ in
$C(x)\cap C(y)$, then it would be a strict descendant of both $x$ and
$y$ by (c), which would imply by Lemma~\ref{lem:strict->anc}
that $x$ and $y$ are connected by a
path, against (a).\smallskip

\noindent \emph{(e)} Any path $r\pathgr i$ contains $h$ and therefore
it contains one of its parents.  But the merge path from $u$ to any
parent of $h$ is a tree path, and hence it must be contained in the subpath
$r\pathgr h$ of $r\pathgr i$.  This implies that $u$ belongs to the
path $r\pathgr i$. 
\smallskip

\noindent \emph{(f)} Let $v$ be a node outside $K$ such that $i\in
C(v)$.  Then, by (e) and Lemma~\ref{lem:strict->anc}, $u$ and $v$ are
connected by a path.  Now, since 
$v\neq h$, $v$ will be an ancestor of one of the parents of $h$, say
$x$.  But then, if $v$ were a descendant of $u$, it would belong to
the only path $u\pathgr x$, which is contained in $K$, against
the assumption that $v$ does not belong to $K$.  Thus, $u$ is a
descendant of $v$.  \smallskip

\noindent \emph{(g)} Let $x$ be an intermediate node of $K$ and assume
that there exists some $w\neq x$ such that $C(w)=C(x)$.  We know by
(b) that $w$ is neither $h$ (because $C(h)\neq C(x)$) nor any
intermediate node of $K$ and therefore, by (f),
$C(u)\subseteq C(w)=C(x)$.  Thus, $C(x)$ contains all clusters of
nodes in $K$, which implies that the merge path not containing $x$
cannot contain any intermediate node (by (d)) and that $x$ is the
child of $u$ in the only merge path of $K$ of length greater than 1
(otherwise, the cluster of its parent in the merge path would strictly contain
$C(x)$, by (b), and would be included in $C(u)$).  \smallskip

\noindent \emph{(h)} Let $v$ and $v'$ be the parents of $h$.  Since
every proper ancestor $w$ of $h$ is an ancestor of $v$ or $v'$, and
hence $C(w)$ contains $C(v)$ or $C(v')$, we deduce that $C(v)$ and
$C(v')$ are the only possible minimal members of $\cIC(N)$ strictly
containing $C(h)$.

Now, if $C(v)$ and $C(v')$ are two different such minimal members of
$\cIC(N)$, then they do not contain each other and therefore $v$ and
$v'$ are not connected by a path.  This implies that neither $v$ nor
$v'$ is the split node $u$ of $K$, and therefore that they are
intermediate in $K$.  Conversely, if only one of these two clusters,
say $C(v)$, is minimal strictly containing $C(h)$, then it is
contained in the other.  By (d), this implies that $v'$ cannot
be intermediate in $K$, and therefore $v'=u$.
\end{proof}

\subsection*{A2\quad Proof of Theorem \ref{prop:RF-AR}}

To ease the task of the reader, we split the proof of Theorem
\ref{prop:RF-AR} into several lemmas.  Throughout this subsection, $N$
stands for a \emph{binary} 1-nested network (or, equivalently, a
binary galled tree) on a fixed set $S$.

\begin{lemma}\label{lem:RF-rR}
The $R_{i;j}$ reduction can be applied to $N$ if, and only if,
$\{i,j\}\in \cIC(N)$ but $\{i\},\{j\}\notin \cIC(N)$.
\end{lemma}

\begin{proof}
If $N$ contains a node $u$ whose children are the tree leaves $i,j$,
then $\{i,j\}=C(u)\in \cIC(N)$, and $\{i\},\{j\}\notin \cIC(N)$
by Lemma \ref{lem:RF-previ}.(a).

Conversely, if $\{i,j\}\in \cIC(N)$ and $\{i\},\{j\}\notin \cIC(N)$,
then $i$ and $j$ are tree leaves (by the binarity of $N$) and their
parents have out-degree greater than 1 by Lemma
\ref{lem:RF-previ}.(a).  Let now $u$ be the parent of $i$.  Since
every internal ancestor of $i$ is an ancestor of its only parent $u$,
the cluster of any internal ancestor of $i$ must contain the cluster
of $u$: in particular, $i\in C(u)\subseteq \{i,j\}$, which implies
(since $\{i\}\notin \cIC(N)$) that $C(u)=\{i,j\}$.  But then, if $i\in
C(v)$ for some internal node $v$, then $\{i,j\}\subseteq C(v)$.  This
shows that every member of $\cIC(N)$ that contains $i$ also contains
$j$.  By symmetry, every member of $\cIC(N)$ that contains $j$ also
contains $i$.  Then, Lemma \ref{lem:RF-previ}.(b) applies.
\end{proof}

\begin{lemma}\label{lem:RF-rG1}
The $G_{i;i_1,\ldots,i_k; \emptyset}$ reduction can be applied to $N$
if, and only if, the following conditions are satisfied:
\begin{itemize}

\item[(1)] $\{i\}\in \cIC(N)$.

\item[(2)] For every $j=2,\ldots,k$, $\{i_j,\ldots,i_k,i\}\in \cC(N)$
with multiplicity 1.

\item[(3)]
$\{i_1,\ldots,i_k,i\}\in \cC(N)$ with multiplicity at least 2.

\item[(4)] Any member of $\cIC(N)$ containing some label among
$i_1,\ldots,i_k$ and not listed in (2)--(3), must contain
$\{i_1,\ldots, i_k, i\}$.
\end{itemize}
\end{lemma}

\begin{proof}
If $N$ contains a reticulation cycle $K$ consisting of the merge paths
$(u,h)$ and $(u,v_1,\ldots,v_k,h)$ (and hence $h$ and $v_1$ are the
only children of $u$), such that the only child of the hybrid node $h$
is the leaf $i$ and the child outside $K$ of each tree node $v_j$ is
the tree leaf $i_j$, then
$$
\begin{array}{l}
C(h)=\{i\}\\
C(v_j)=\{i_j,\ldots,i_k,i\},\quad j=1,\ldots,k\\
C(u)=\{i_1,\ldots,i_k,i\}
\end{array}
$$
and hence $\cIC(N)$ contains all clusters listed in (1)--(2), the
latter with multiplicity 1 by Lemma \ref{lem:1n-cicles}.(g), as well
as the cluster given in (3) with multiplicity at least 2.  Now, let
$v$ be any internal node of $N$ not belonging to $K$ and such that
$C(v)$ contains some label $i_1,\ldots,i_k$.  If $i_j\in C(v)$, then
$i_j$'s only parent $v_j$ must also be a descendant of $v$.  But then
$i\in C(v_j)\subseteq C(v)$ implies that $C(u)\subseteq C(v)$ by
Lemma \ref{lem:1n-cicles}.(f), as (4) claims.

Conversely, assume that (1)--(4) are satisfied.  Then, the parent $h$
of $i$ has out-degree 1 and therefore it is hybrid, and, by Lemma
\ref{lem:1n-cicles}.(h), its parents are connected by a path, because
there is only one minimal element of $\cIC(N)$ strictly containing
$\{i\}$, namely $\{i_k,i\}$.  Therefore, the reticulation cycle $K$
for $h$ consists of an arc $(u,h)$ and a tree path
$(u,v_1,\ldots,v_l,h)$ with $l\geq 1$.  In this situation, Lemma
\ref{lem:1n-cicles} implies that:
\begin{itemize}
\item $C(v_l)$ is the minimal element of $\cIC(N)$ strictly containing
$\{i\}$;

\item $C(v_l)\subsetneq C(v_{l-1}) \subsetneq \cdots \subsetneq
C(v_1)$, and then, by Lemma \ref{lem:RF-previ}.(c), each $C(v_j)$,
$j=1,\ldots,l-1$, is the minimal element of $\cIC(N)$ containing
$C(v_{j+1})$;

\item $C(v_2),\ldots,C(v_l)$ appear with multiplicity 1 in $\cIC(N)$;

\item $C(u)=C(v_1)$, because the only children of $u$ are $v_1$ and
$h$.
\end{itemize}
On the other hand, (1)--(4) imply that
\begin{itemize}
\item The minimal element of $\cIC(N)$ strictly containing $\{i\}$ is
$\{i_k,i\}$;

\item $\{i_k,i\}\subsetneq \{i_{k-1},i_k,i\}\subsetneq\ldots
\subsetneq \{i_1,\ldots,i_k,i\}$, and each $\{i_j,\ldots,i_k,i\}$,
$j=1,\ldots,k-1$, is the minimal element of $\cIC(N)$ strictly
containing $\{i_{j+1},\ldots,i_k,i\}$;

\item $\{i_2,\ldots,i_k,i\},\ldots, \{i_k,i\}$ appear with
multiplicity 1 in $\cIC(N)$;

\item $ \{i_1,\ldots,i_k,i\}$ appears with multiplicity at least 2 in
$\cIC(N)$.
\end{itemize}
The only possibility of making these two lists of properties
compatible is that $k=l$ and $C(v_j)=\{i_j,\ldots,i_k,i\}$ for every
$i=1,\ldots,k$.

It remains to prove that the only child of every $v_j$ outside $K$ is
the corresponding leaf $i_j$.  Let $w_j$ be the only parent of $i_j$;
we want to prove that $w_j=v_j$.  Since $i_j\in C(v_j)$, there exists
a path $v_j\pathgr w_j$ and hence $C(w_j)\subseteq C(v_j)$.  On the
other hand, $i_j\in C(w_j)$ implies, by (4), that $i\in C(w_j)$ and
therefore, by Lemma \ref{lem:1n-cicles}.(f), either $w_j$ belongs to
$K$ or it is an ancestor of $u$.  The second case cannot hold, because
$w_j$ is a proper descendant of $u$.  Therefore, $w_j$ is a node of
$K$ that is a descendant of $v_j$ and an ancestor of $i_j$: it must be
$v_j$.  \end{proof}

\begin{lemma}\label{lem:RF-rG2}
The $G_{i;i_1,\ldots,i_k; i'_1,\ldots,i'_{k'}}$ reduction, with $k\geq
k'>0$, can be applied to $N$ if, and only if, the following conditions
are satisfied:
\begin{itemize}

\item[(1)] $\{i\}\in \cIC(N)$.
 
\item[(2)] For every $j=1,\ldots, k$, $\{i_j,\ldots,i_k,i\}\in \cC(N)$
with multiplicity 1.

\item[(3)] For every $j=1,\ldots, k'$, $\{i'_j,\ldots,i'_{k'},i\}\in
\cC(N)$ with multiplicity 1.

\item[(4)]
$\{i_1,\ldots,i_{k-1},i_k,i'_1,\ldots,i'_{k'-1}, i'_{k'}, i\}\in
\cC(N)$.

\item[(5)] Any member of $\cIC(N)$ containing some label among
$i_1,\ldots,i_k,i'_1,\ldots,i'_{k'}$ and not listed in (1)--(4), must
contain $\{i_1,\ldots,i_k,i'_1,\ldots,i'_{k'},i\}$.
\end{itemize}
\end{lemma}

\begin{proof}
The proof that if $N$ contains a reticulation cycle $K$ consisting of
the merge paths $(u,v_1,\ldots,v_k,h)$ and
$(u,v'_1,\ldots,v'_{k'},h)$, with $k\geq k'>0$, such that the only
child of the hybrid node $h$ is the leaf $i$, the child outside $K$ of
each tree node $v_j$ is the tree leaf $i_j$, and the child of each
tree node $v'_j$ outside $K$ is the tree leaf $i'_j$, then it
satisfies conditions (1) to (5), is similar to the proof of the
corresponding implication in the previous lemma, and we do not repeat
it here.

As far as the converse implication goes, assume that conditions
(1)--(5) in the statement are satisfied.  Then, the parent $h$ of $i$
has out-degree 1 and therefore it is hybrid, and, by Lemma
\ref{lem:1n-cicles}.(h), its two parents are not connected by a path,
because there are two minimal elements of $\cIC(N)$ strictly containing
$\{i\}$, namely $\{i_k,i\}$ and $\{i'_{k'},i\}$.  Therefore, the
reticulation cycle $K$ for $h$ consists of two merge paths
$(u,v_1,\ldots,v_l,h)$ and $(u,v'_1,\ldots,v'_{l'},h)$ with $l,l'\geq
1$.  In this situation, Lemma \ref{lem:1n-cicles} implies that:
\begin{itemize}
\item $C(v_l)$ and $C(v'_{l'})$ are the minimal elements of $\cIC(N)$
strictly containing $\{i\}$;

\item $C(v_l)\subsetneq \cdots \subsetneq C(v_1)$, and then, by Lemma
\ref{lem:RF-previ}.(c), each $C(v_j)$, $j=1,\ldots,l-1$, is the
minimal element of $\cIC(N)$ containing $C(v_{j+1})$;

\item $C(v'_{l'})\subsetneq \cdots \subsetneq C(v'_1)$, and then, by
Lemma \ref{lem:RF-previ}.(c), each $C(v'_j)$, $j=1,\ldots,l'-1$, is
the minimal element of $\cIC(N)$ containing $C(v'_{j+1})$;

\item $C(v_1),\ldots,C(v_l),C(v'_1),\ldots, C(v'_{l'})$ appear with
multiplicity 1 in $\cIC(N)$;

\item the minimal element of $\cIC(N)$ strictly containing $C(v_1)$ is
the same as the minimal element of $\cIC(N)$ strictly containing
$C(v'_1)$, and it is $C(u)$.
\end{itemize}
On the other hand, (1)--(5) imply that:
\begin{itemize}
\item The minimal elements of $\cIC(N)$ strictly containing $\{i\}$
are $\{i_k,i\}$ and $\{i'_{k'},i\}$;

\item $\{i_k,i\}\subsetneq\ldots \subsetneq \{i_1,\ldots,i_k,i\}$, and
each $\{i_j,\ldots,i_k,i\}$, $j=1,\ldots,k-1$, is the minimal element
of $\cIC(N)$ strictly containing $\{i_{j+1},\ldots,i_k,i\}$;

\item $\{i'_k,i\}\subsetneq\ldots \subsetneq
\{i'_1,\ldots,i'_{k'},i\}$, and each $\{i'_j,\ldots,i'_{k'},i\}$,
$j=1,\ldots,k'-1$, is the minimal element of $\cIC(N)$ strictly
containing $\{i'_{j+1},\ldots,i'_{k'},i\}$;

\item $\{i_k,i\},\ldots, \{i_1,\ldots,i_k,i\}, \{i'_k,i\},\ldots,
\{i'_1,\ldots,i'_{k'},i\}$ appear with multiplicity 1 in $\cIC(N)$;

\item the minimal element of $\cIC(N)$ strictly containing $
\{i_1,\ldots,i_{k},i\}$ is the same as the minimal element containing
$ \{i'_1,\ldots,i'_{k'},i\}$, and it is $
\{i_1,\ldots,i_{k},i'_1,\ldots,i'_{k'},i\}$.

\end{itemize}
The only possibility of making these two lists of properties
compatible is that (up to the interchange of $k$ and $k'$) $k=l$,
$k'=l'$, $C(v_j)=\{i_j,\ldots,i_k,i\}$ for every $j=1,\ldots,k$, and
$C(v'_j)=\{i'_j,\ldots,i'_k,i\}$ for every $j=1,\ldots,k'$.

It remains to prove that the only child of every $v_j$ (respectively  $v'_j$)
not belonging to the reticulation cycle for $h$ is the corresponding
leaf $i_j$ (respectively  $i'_j$).  This fact can be proved using the same
argument as in the last paragraph of the proof of the previous lemma.
\end{proof}

Lemmas \ref{lem:RF-rR} to \ref{lem:RF-rG2} prove that the fact
that a given $R$ or $G$ reduction can be applied to $N$ only depends
on $\cC(N)$, from where point (A) in Theorem \ref{prop:RF-AR}
follows.  As far as point (R) goes, it is a consequence of the
following straightforward lemma that shows that the application of a
specific $R$ or $T$ reduction to $N$ affects $\cC(N)$ in a way that
does not depend on $N$ itself, but only on its cluster
representation; we leave its easy proof to the reader.

\begin{lemma}\label{lem:RF-R}
\begin{enumerate}[(a)]

\item If the $R_{i;j}$ reduction can be applied to $N$, then
$\cC(R_{i;j}(N))$ is obtained by removing from $\cC(N)$ the clusters
$\{i\}$ and $\{j\}$, and them removing from all remaining clusters the
label $j$.

\item If the $G_{i;i_1,\ldots,i_k; \emptyset}$ reduction can be
applied to $N$, then $\cC(G_{i;i_1,\ldots,i_k; \emptyset}(N))$ is
obtained by first removing from $\cC(N)$ all clusters listed in points
(1)--(2) of Lemma \ref{lem:RF-rG1}, one copy of the cluster given in
point (3) therein, and the clusters $\{i_2\},\ldots, \{i_k\}$, and
then removing the labels $i_2,\ldots,i_k$ from all remaining clusters.

\item If the $G_{i;i_1,\ldots,i_k; i'_1,\ldots,i'_{k'}}$ (with $k'\neq
0$) reduction can be applied to $N$, then $\cC(G_{i;i_1,\ldots,i_k;
i'_1,\ldots,i'_{k'}}(N))$ is obtained by first removing from $\cC(N)$
all clusters listed in points (1)--(3) of Lemma \ref{lem:RF-rG2} and
the clusters $\{i_2\},\ldots, \{i_k\},\{i'_1\},\ldots,\{i'_{k'}\}$,
and then removing the labels $i_2,\ldots,i_k,i'_1,\ldots,i'_{k'}$ from
all remaining clusters.  \qed
\end{enumerate}
\end{lemma}

\subsection*{A3\quad Proof of Theorem \ref{prop:tri-AR}}

As in the previous subsection, we split the proof of Theorem
\ref{prop:tri-AR} into several lemmas to increase its readability.  In
the rest of this subsection, $N$ stands for an arbitrary 1-nested
network on some given set $S$.  Since the set $S$ is fixed, for
every node $v$ of $N$, if $A(v)=\{i_1,\ldots,i_k\}$ and
$B(v)=\{j_1,\ldots,j_l\}$, we shall use the following notation to
denote the tripartition $\theta(v)$:
$$
\theta(v)=\{i_1,\ldots,i_k\mid j_1,\ldots,j_l\}.
$$
  To simplify the notations, we shall denote by $\cIT(N)$ the multiset
  of tripartitions of its internal nodes, which is obtained by
  removing from $\theta(N)$ one copy of every tripartition $\{i\mid
  \emptyset\}$ with $i\in S$.

\begin{lemma}\label{lemma:tri-germans}
Two leaves $i,j$ are tree leaves and siblings if, and only if, the
following conditions are satisfied:
\begin{itemize}

\item[(a)] There exists an internal node $v$ such that $i,j\in A(v)$
and $C(v)$ is contained in the cluster of any internal ancestor of $i$
or $j$.

\item[(b)] For every node $w$ of $N$ such that $i,j\in C(w)$, it
happens that either $i,j\in A(w)$ or $i,j\in B(w)$.
\end{itemize}
Moreover, when $i$ and $j$ are sibling tree leaves, they are the only
children of their parent if, and only if, the node $v$ in point (a) is
such that $C(v)=\{i,j\}$.
\end{lemma}

\begin{proof}
 If $i,j$ are two sibling tree leaves and $v$ is their common parent,
 then they are strict descendants of $v$ and $C(v)$ is contained in
 the cluster of any ancestor of $i$ or $j$.  Let now $w$ be any node
 such that $i,j\in C(w)$.  Then, $w$ is ancestor of $v$.  If $v$ is a
 strict descendant of $w$, then $i,j$ are also strict descendants of
 $w$, and if $v$ is a non-strict descendant of $w$, then $i,j$ are
 also non-strict descendants of $w$.  Therefore, either $i,j\in A(w)$
 or $i,j\in B(w)$.  This finishes the proof of the `only if'
 implication.

As far as the converse implication goes, the existence of the internal
node $v$ with $i,j\in A(v)$ and such that $C(v)$ is contained in the
cluster of every ancestor of $i$ or $j$ implies that there does not
exist any internal node whose cluster contains one of the labels $i,j$
but not the other, and therefore, by Lemma \ref{lem:RF-previ}.(b),
that $i$ and $j$ are siblings.  Let $v_0$ be a common parent of
them: then, on the one hand, $i,j\in C(v_0)$ implies that
$C(v)\subseteq C(v_0)$, and, on the other hand, since $i,j\in A(v)$,
$v_0$ must be a descendant of $v$, and therefore $C(v_0)\subseteq
C(v)$.  We conclude that $C(v_0)=C(v)$.

Let us prove now that $i,j\in A(v_0)$.  Indeed, if one of them
were a non-strict descendant of $v_0$, then by (b) both would be
non-strict descendants of it.  By Lemma \ref{lem:RF-previ}.(d), and
taking into account that $v_0$ is a parent of $i$ and $j$, this would
imply that $i$ and $j$ are hybrid leaves and $v_0$ intermediate in
their reticulation cycles, which would contradict the 1-nested
condition.

This implies that there would exist paths from the root of $N$ to $i$
and $j$ that do not contain $v_0$.  This could only happen if both $i$
and $j$ were hybrid leaves and $v_0$ intermediate in their
reticulation cycles (if it were the split node of one of them, the
corresponding hybrid leaf would be a strict descendant of it by Lemma
\ref{lem:1n-cicles}.(e)), Let us prove now that $i$ and $j$ are tree
leaves.  Indeed, if, say, $i$ is a hybrid leaf and $v_0'$ its other
parent, then, since $i\in A(v_0)$, $v_0'$ is a descendant of $v_0$ and
then intermediate in the reticulation cycle for $i$ (which would have
$v_0$ as split node).  Now, since $i\in C(v_0')$, it must happen that
$j\in C(v_0')$ and, since $v_0'$ cannot be an ancestor of $v_0$, we
conclude that $j$ is also hybrid and that $v_0'$ is an ancestor of its
other parent.  But then, $v_0'$ is also intermediate in the
reticulation cycle for $j$ (which consists of the arc $(v_0,j)$ and
the merge path $v_0\pathgr v_0'\pathgr j$), which is impossible.  This
shows that $i$ and, by symmetry, $j$ are tree leaves.

This finishes the proof that $i$ and $j$ are tree sibling leaves if,
and only if, (a) and (b) are satisfied; moreover, from this proof we
deduce that we can take as $v$ in (a) the common parent of $i$ and
$j$.  Now, as far as the last assertion in the statement, if $i$ and
$j$ are the only children of their common parent $v$, it is clear that
$C(v)=\{i,j\}$.  Conversely, if $v$ has a child $u$ different from $i$
and $j$, then $u$ cannot be an ancestor of $i$ and $j$, and therefore
any descendant leaf of it is an element of $C(v)$ different from $i$
and $j$, which shows that $\{i,j\}\subsetneq C(v)$.
  \end{proof}

As a direct consequence of this lemma we obtain the following two
results.

\begin{lemma}\label{lem:tri-rR}
The $R_{i;j}$ reduction can be applied to $N$ if, and only if, the
following conditions are satisfied:
\begin{itemize}

\item[(1)] There exists an internal node $v$ such that
$\theta(v)=\{i,j\mid \emptyset\}$ and $C(v)$ is contained in the
cluster of any internal ancestor of $i$ or $j$.

\item[(2)] For every node $w$ of $N$ such that $i,j\in C(w)$, it
happens either that $i,j\in A(w)$ or $i,j\in B(w)$.  \qed
\end{itemize}
\end{lemma}

\begin{lemma}\label{lem:tri-rT}
The $T_{i;j}$ reduction can be applied to $N$ if, and only if, the
following conditions are satisfied:
\begin{itemize}

\item[(1)] There exists an internal node $v$ such that $i,j\in A(v)$,
$\{i,j\}\subsetneq C(v)$, and $C(v)$ is contained in the cluster of
any other internal ancestor of $i$ or $j$.

\item[(2)] For every node $w$ of $N$ such that $i,j\in C(w)$, it
happens either that $i,j\in A(w)$ or $i,j\in B(w)$.  \qed
\end{itemize}
\end{lemma}

Let us consider now the $G$ and $\ovG$ reductions.  In contrast to the
corresponding lemmas in \S A2, here we do not need to distinguish
between $k'=0$ and $k'>0$.

\begin{lemma}\label{lem:tri-rG1}
The $G_{i;i_1,\ldots,i_k; i'_1,\ldots,i'_{k'}}$ reduction (with $k\geq
k'\geq 0$) can be applied to $N$ if, and only if, the following
conditions are satisfied:
\begin{itemize}

\item[(1)] $\{i\mid\emptyset \}\in \cIT(N)$.

\item[(2)] For every $j=1,\ldots, k$, $\{i_j,\ldots,i_{k}\mid i\}\in
\theta(N)$ with multiplicity 1.

\item[(3)] For every $j=1,\ldots, k'$, $\{i'_j,\ldots,i'_{k'}\mid
i\}\in \theta(N)$ with multiplicity 1.

\item[(4)] For every $\theta(v)\in \cIT(N)$, if $C(v)$ contains some
label among $i_1,\ldots,i_k,i'_1,\ldots,i'_{k'}$ and $\theta(v)$ is
not listed in (2) or (3), then either
$i,i_1,\ldots,i_k,i'_1,\ldots,i'_{k'}\in A(v)$ or
$i,i_1,\ldots,i_k,i'_1,\ldots,i'_{k'}\in B(v)$.
\end{itemize}
\end{lemma}

\begin{proof}
 If $N$ contains a reticulation cycle $K$ consisting of the merge
 paths $(u,v_1,\ldots,v_k,h)$ and $(u,v'_1,\ldots,v'_{k'},h)$ such
 that the only child of the hybrid node $h$ is the tree leaf $i$ and
 each tree node $v_j$ (respectively  $v'_j$) has only one child outside $K$
 and it is the tree leaf $i_j$ (respectively  $i'_j$), then
$$
\begin{array}{l}
\theta(h)=\{i\mid \emptyset\}\\
\theta(v_j)=\{i_j,\ldots,i_k\mid i\},\quad j=1,\ldots,k\\
 \theta(v'_j)=\{i'_j,\ldots,i'_{k'}\mid i\},\quad j=1,\ldots,k'
\end{array}
$$
and hence $\cIT(N)$ contains all tripartitions listed in points
(1)--(3).  Let now $v$ be any internal node different from
$v_1,\ldots,v_k,v'_1,\ldots,v'_{k'}$ that is an ancestor of some leaf
$i_1,\ldots, i_k,i'_1,\ldots,i'_{k'}$, say that $i_j\in C(v)$.  Then
it will be an ancestor of its parent $v_j$ and in particular $i\in
C(v_j)\subseteq C(v)$.  Then, by Lemma \ref{lem:1n-cicles}.(f), $u$ is
a descendant of $v$.  Now, if $u$ is a strict descendant of $v$, then
the leaves $i_1,\ldots, i_k,i'_1,\ldots,i'_{k'},i$ are also strict
descendants of $v$, while if $u$ is a non-strict descendant of $v$,
then they are also non-strict descendants of $v$.  This proves (4) and
that all tripartitions listed in (2) and (3) appear only once in
$\theta(N)$ (Lemma \ref{lem:1n-cicles}.(g) did not guarantee it for
$\theta(v_1)$ when $k'=0$).  This finishes the proof of the `only if'
implication.

Conversely, assume that (1)--(4) are satisfied.  Using only the
information of the clusters and arguing as in the proof of the `if'
implication in Lemmas \ref{lem:RF-rG1} and \ref{lem:RF-rG2}, we
already deduce that the parent $h$ of $i$ is hybrid, it has out-degree
1 and $i$ is a tree child of it (using Lemma \ref{lem:RF-previ}.(a)),
that the reticulation cycle $K$ for $h$ consists of two tree merge
paths $(u,v_1,\ldots,v_l,h)$ and $(u,v'_1,\ldots,v'_{l'},h)$, with
$l\geq k$ and $l'\geq k'$, and that
$\theta(v_{l-j})=\{i_{k-j},\ldots,i_k\mid i\}$, for every
$j=0,\ldots,k-1$, and
$\theta(v'_{l'-j})=\{i'_{k'-j},\ldots,i'_{k'}\mid i\}$ for every
$j=0,\ldots,k'-1$.  Now, if $l>k$, $C(v_{l-k})$ would strictly contain
$C(v_{l-(k-1)})$ and $i_1$ would be a strict descendant of $ v_{l-k}$
(because it is a strict descendant of its tree child $v_{l-(k-1)}$)
but $i$ would be a non-strict descendant of it, which would contradict
(4).  This implies then that $k=l$ and $C(v_j)=\{i_j,\ldots,i_k,i\}$
for every $i=1,\ldots,k$, and then, by symmetry, $k'=l'$ and
$C(v'_j)=\{i'_j,\ldots,i'_{k'},i\}$ for every $i=1,\ldots,k'$.

It remains to prove that the only child of every $v_j$ (respectively, $v'_j$)
not belonging to the reticulation cycle for $h$ is the corresponding
tree leaf $i_j$ (respectively, $i'_j$).  Let us prove first that each $i_j$
is a tree leaf.  Indeed, if $i_j$ were hybrid, then, since $i_j\in
A(v_j)$, $v_j$ would be an ancestor of the split node $w_j$ of the
reticulation cycle for $i_j$.  Since $i_j$ only belongs to the
clusters of the nodes in $K$ that are ancestors of $v_j$, we conclude
that $w_j$ does not belong to $K$ and then, since $i_j\in A(w_j)$, by
(4) we have that $i\in A(w_j)$ and hence, by Lemma
\ref{lem:1n-cicles}.(f), that $w_j$ is a strict ancestor of $u$, which
is impossible.

Let now $u_j$ be a child of $v_j$ different from $i_j$ and from
$v_j$'s child in $K$.  This node must be internal, because the other
descendant leaves $i_{j+1},\ldots,i_k,i$ of $v_j$ are tree leaves and
descendants of proper descendants of $v_j$ in $K$, and therefore $v_j$
is not their parent.  Then, since $C(u_j)\subseteq
C(v_j)=\{i_j,\ldots,i_k,i\}$, by (4) we conclude that $i\in C(u_j)$
and hence, since $u_j$ does not belong to $K$, by Lemma
\ref{lem:1n-cicles}.(f) we conclude that $u_j$ is an ancestor of $u$,
which is impossible.  This shows that $v_j$ does not have any child
outside $K$ different from $i_j$, and moreover, since $i_j$ is a
descendant of $v_j$, that it is its child.
\end{proof}

\begin{lemma}\label{lem:tri-rovG1}
The $\ovG_{i;i_1,\ldots,i_k; i'_1,\ldots,i'_{k'}}$ reduction (with
$k\geq k'\geq 0$) can be applied to $N$ if, and only if, the following
conditions are satisfied:
\begin{itemize}

\item[(1)] $\{i\}\notin \cIC(N)$.

\item[(2)] For every $j=1,\ldots, k$, $\{i_j,\ldots,i_{k}\mid i\}\in
\theta(N)$ with multiplicity 1.

\item[(3)] For every $j=1,\ldots, k'$, $\{i'_j,\ldots,i'_{k'}\mid
i\}\in \theta(N)$ with multiplicity 1.

\item[(4)] For every $\theta(v)\in \cIT(N)$, if $C(v)$ contains some
label among $i_1,\ldots,i_k,i'_1,\ldots,i'_{k'}$ and $\theta(v)$ is
not listed in (2) or (3), then either
$i,i_1,\ldots,i_k,i'_1,\ldots,i'_{k'}\in A(v)$ or
$i,i_1,\ldots,i_k,i'_1,\ldots,i'_{k'}\in B(v)$.
\end{itemize}
\end{lemma}

\begin{proof}
If $N$ contains a reticulation cycle $K$ consisting of the merge paths
$(u,v_1,\ldots,v_k,h)$ and $(u,v'_1,\ldots,v'_{k'},h)$ such that the
hybrid node $h$ is the leaf $i$ and each tree node $v_j$ (respectively,
$v'_j$) has only one child outside $K$ and it is the tree leaf $i_j$
(respectively  $i'_j$), then, by Lemma \ref{lem:RF-previ}.(a), $\{i\}\notin
\cIC(N)$, and
$$
\begin{array}{l}
\theta(v_j)=\{i_j,\ldots,i_k\mid i\},\quad j=1,\ldots,k\\
 \theta(v'_j)=\{i'_j,\ldots,i'_{k'}\mid i\},\quad j=1,\ldots,k'\\
\end{array}
$$
and hence $N$ satisfies (1)--(3).  The rest of the `only if'
implication can be proved as in Lemma \ref{lem:tri-rG1}.

Conversely, assume that (1)--(4) are satisfied.  To begin with, let us
prove that $i$ is a hybrid leaf.  Indeed, if it were a tree leaf, then
its parent $v$ would be a strict ancestor of $i$, and therefore
$\theta(v)$ would be none of the tripartitions listed in (2) or (3).
On the other hand, $v$ would be a descendant of the node $w$ having
tripartition $\{i_k\mid i\}$, which would imply, since $\{i\}\neq
C(v)$ by (1), that $i_k\in C(v)$.  Then, by (4) and since $i\in A(v)$,
$i_k$ would also be a strict descendant of $v$ .  This would imply
that $w$ is a strict ancestor of $v$: any path $r\pathgr v$ not
containing $w$ followed by a path $v\pathgr i_k$ (that does not
contain $w$ because $w$ is an ancestor of $v$) would form a path
$r\pathgr i_k$ not containing $w$, against the assumption that $i_k\in
A(w)$.  But then the tree child $i$ of $v$ would be also a strict
descendant of $w$, which would contradict the assumption that $i\in
B(w)$.

Let us also denote by $h$ this hybrid leaf labeled with $i$, so that
$C(h)=\{i\}$.  Since we can still apply Lemma \ref{lem:1n-cicles}, the
same argument as in the proof of the `if' implication in Lemma
\ref{lem:tri-rG1} implies that the reticulation cycle $K$ for $h$
consists of two merge paths $(u,v_1,\ldots,v_k,h)$ and
$(u,v'_1,\ldots,v'_{k'},h)$ such that $C(v_j)=\{i_j,\ldots,i_k,i\}$,
for every $j=1,\ldots,k$, and $C(v'_j)=\{i'_j,\ldots,i'_{k'},i\}$ for
every $j=1,\ldots,k'$.

The proof that the only child of every $v_j$ (respectively, $v'_j$) not
belonging to $K$ is the corresponding tree leaf $i_j$ (respectively, $i'_j$)
is also similar to the one given for the corresponding fact in Lemma
\ref{lem:tri-rG1}, and we do not repeat it here.  \end{proof}

Lemmas \ref{lem:tri-rR} to \ref{lem:tri-rovG1} prove condition (A) in
Lemma \ref{lem:basic-met} for the $R$, $T$, $G$, and $\ovG$ reductions
and the tripartitions representation.  As far as point (R) goes, it is
a consequence of the following lemma.

\begin{lemma}\label{lem:tri-R}
\begin{enumerate}[(a)]

\item If the $R_{i;j}$ reduction can be applied to $N$, then
$\theta(R_{i;j}(N))$ is obtained by removing the tripartitions
$\{i\mid\emptyset\}$ and $\{j\mid\emptyset\}$ from $\theta(N)$, and
then removing the label $j$ from all remaining tripartitions in
$\theta(N)$.

\item If the $T_{i;j}$ reduction can be applied to $N$, then
$\theta(T_{i;j}(N))$ is obtained by removing the label $j$ from all
tripartitions in $\theta(N)$.

\item If the $G_{i;i_1,\ldots,i_k; i'_1,\ldots,i'_{k'}}$ reduction can
be applied to $N$, then $\theta(G_{i;i_1,\ldots,i_k;
i'_1,\ldots,i'_{k'}}(N))$ is obtained by first removing from
$\theta(N)$ all tripartitions listed in points (1)--(3) of Lemma
\ref{lem:tri-rG1} and the tripartitions $\{i_2\mid\emptyset\},\ldots,
\{i_k\mid\emptyset\},\{i'_1\mid\emptyset\},\ldots,
\{i'_{k'}\mid\emptyset\}$, and then removing the labels
$i_2,\ldots,i_k,i'_1,\ldots,i'_{k'}$ from all remaining tripartitions.

\item If the $\ovG_{i;i_1,\ldots,i_k; i'_1,\ldots,i'_{k'}}$ reduction
can be applied to $N$, then $\theta(\ovG_{i;i_1,\ldots,i_k;
i'_1,\ldots,i'_{k'}}(N)$ is obtained by first removing from
$\theta(N)$ all tripartitions listed in point (2) of Lemma
\ref{lem:tri-rovG1} and the tripartitions
$\{i_2\mid\emptyset\},\ldots,
\{i_k\mid\emptyset\},\{i'_1\mid\emptyset\},\ldots,
\{i'_{k'}\mid\emptyset\}$, and then removing the labels
$i_2,\ldots,i_k,i'_1,\ldots,i'_{k'}$ from all remaining tripartitions.
\qed
\end{enumerate}
\end{lemma}

\subsection*{A4\quad Proof of Theorem \ref{prop:snod-AR}}

We also split the proof of Theorem \ref{prop:snod-AR} into several
lemmas, in parallel to the preceding subsections.  In the rest of this
subsection, $N$ stands for a \emph{semibinary} 1-nested network on
$S=\{1,\ldots,n\}$.  Notice that all leaves in $N$ are of tree type.

For every pair of nodes $u,v$ in $N$, we shall denote by $CA(u,v)$ the
set of common ancestors of $u$ and $v$.  By Lemma \ref{lem:lcsa-1n},
$[u,v]$ is the element of $CA(u,v)$ that is a descendant of all other
nodes in this set.

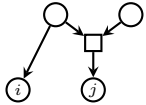
\begin{figure}[htb]
\centering
  \begin{tikzpicture}[thick,>=stealth,scale=0.25]
    \draw(0,0) node[tre] (x) {};
    \draw(-2,-4) node[tre] (i) {};\etq i
    \draw(2,-1.5) node[hyb] (y) {};
    \draw(2,-4) node[tre] (j) {};\etq j
    \draw(4,0) node[tre] (z) {};
\draw[->] (z)--(y);
    \draw[->](x)--(i);
    \draw[->](x)--(y);
    \draw[->](y)--(j);
  \end{tikzpicture}
\caption{\label{fig:qsibl} 
$j$ is a quasi-sibling of $i$.}
\end{figure}

The following result summarizes what Lem.~5 and Cor.~4 in
\cite{cardona.ea:tcbb:comparison.2:2008} say about $N$.  Although
these results were stated therein for tree-child time consistent
evolutionary networks with out-degree 1 hybrid nodes, it is
straightforward to check that the time consistency is not used
anywhere in their proofs, and therefore their thesis also holds for
tree-child (and, in particular, for 1-nested) semibinary hybridization
networks.  In the following statement, and henceforth, by saying that
a leaf $j$ is a \emph{quasi-sibling} of a leaf $i$, we mean that the
parent of $j$ is a hybrid node that is a sibling of $i$:
cf.~Fig.~\ref{fig:qsibl}.

\begin{lemma}\label{lem:importat}
\label{lem:ell=1}
Let $i,j$ be any labels in $S$.
\begin{enumerate}[(a)]
\item $\ell_{N}(i,j)=1$ if, and only if, the parent of $i$ is an
ancestor of $j$.

\item The leaves $i,j$ are siblings if, and only if,
$\ell_N(i,j)=\ell_N(j,i)=1$.

\item The leaf $j$ is a quasi-sibling of the leaf $i$ if, and only if,
$\ell_N(i,j)=1$, $\ell_N(j,i)=2$, and $\ell_N(j,k)>1$ for every $k\in
S\setminus\{i,j\}$.\qed
\end{enumerate}
\end{lemma}

As a consequence of this lemma, we have the following results.

\begin{lemma}\label{lem:nod-rR}
The $R_{i;j}$ reduction can be applied to $N$ if, and only if,
$\ell_N(i,j)=\ell_N(j,i)=1$ and $\ell_N(i,k)>1$ for every $k\in
S\setminus\{i,j\}$.
\end{lemma}

\begin{proof}
The $R_{i;j}$ reduction can be applied to $N$ if, and only if, $i,j$
are sibling leaves and their parent has out-degree 2.  By the previous
lemma we already know that $\ell_N(i,j)=\ell_N(j,i)=1$ if, and only
if, $i,j$ are sibling leaves.  Thus, it only remains to prove that the
parent of $i$ and $j$ has out-degree $2$ if, and only if,
$\ell_N(i,k)>1$ for every $k\in S\setminus\{i,j\}$.  Now, if
there is a leaf $k\neq j$ such that $\ell_N(i,k)=1$, then the parent
of $i$ and $j$ is also an ancestor of $k$, which means that it has
out-degree at least $3$.  Conversely, if the parent of $i$ and $j$ has
out-degree at least $3$ and $v$ is a child of it other than $i,j$,
then $\ell_N(i,k)=1$ for every descendant leaf $k$ of $v$.
\end{proof}

A similar argument, using that the $T_{i;j}$ reduction can be applied
to $N$ if, and only if, $i$ and $j$ are tree sibling leaves and their
parent has some other child, proves the following result.

\begin{lemma}\label{lem:nod-rT}
The $T_{i;j}$ reduction can be applied to $N$ if, and only if,
$\ell_N(i,j)=\ell_N(j,i)=1$ and there exists some $k\in
S\setminus\{i,j\}$ such that $\ell_N(i,k)=1$.\qed
\end{lemma}

We have now the following lemmas for the $G$ reductions.
%



\begin{lemma}\label{lem:nod-rG1}
The $G_{i;i_1,\ldots,i_k;\emptyset}$ reduction can be applied to $N$
if, and only if, the following conditions are satisfied:
\begin{itemize}
\item[(1)] $\ell_N(i,l)>1$ for every $l\in S\setminus\{i\}$.

\item[(2)] $\ell_N(i_k,i)=1$ and $\ell_N(i,i_k)=2$.

\item[(3)] For every $j=1,\ldots,k-1$, $\ell_N(i_j,i_{j+1})=1$ and
$\ell_N(i_{j+1},i_j)=2$.

\item[(4)] For every $j=1,\ldots,k$ and for every $l\neq
i_{j},...,i_k,i$, $\ell_N(i_j,l)>1$.

\item[(5)] For every $j=1,\ldots,k$, $\ell_N(i,i_j)=k-j+2$.

\item[(6)] For every $l\notin \{i,i_1,...,i_k\}$,
$\ell_N(i,l)=\ell_N(i_1,l)$ and $\ell_N(l,i)=\ell_N(l,i_1)$.

\end{itemize}
\end{lemma}
 
\begin{proof}
Assume that $N$ contains a reticulation cycle $K$ consisting of the
merge paths $(u,v_1,\ldots,v_k,h)$ and $(u,h)$ such that the only
child of the hybrid node $h$ is the leaf $i$ and each tree node $v_j$
has only one child outside $K$, and it is the tree leaf $i_j$.  Then,
(1) and (2) are satisfied because $i$ is a quasi-sibling of $i_k$, (3)
is satisfied because the parent $v_{j+1}$ of each $i_{j+1}$ is a
sibling of $i_j$, (4) is satisfied because the only descendant leaves
of the parent $v_j$ of $i_j$ are $i_j,i_{j+1},\ldots,i_k,i$, and (5)
is satisfied because $[i_j,i]=v_j$ and the only path $v_j\pathgr i$
has length $k-j+2$.  As far as condition (6) goes, let $l$ be any
label different from $i,i_1,\ldots,i_k$.  Then, $l$ is not a
descendant of $v_1$ and therefore every common ancestor of $i$ or
$i_1$ and $l$ must be an ancestor of $u$.  This implies that
$CA(i,l)=CA(u,l)=CA(i_1,l)$, from where we deduce that
$[i,l]=[u,l]=[i_1,l]$.  This clearly implies that
$\ell_N(l,i)=\ell_N(l,i_1)$.  On the other hand, any shortest path
$[u,l]\pathgr i$ will consist of a shortest path $[u,l]\pathgr u$
followed by the path $(u,h,i)$, and any shortest path $[u,l]\pathgr
i_1$ will consist of a shortest path $[u,l]\pathgr u$ followed by the
path $(u,v_1,i_1)$, which implies that $\ell_N(i,l)=\ell_N(i_1,l)$.

Conversely, assume that $N$ satisfies conditions (1) to (6).  Then,
conditions (1) and (2) imply that $i$ is a quasi-sibling of $i_k$: let
$h$ be the hybrid parent of $i$ and let $v_k$ be the parent of $i_k$
and $h$, which will be a tree node because it has out-degree at least
2.  Now, condition (3) implies that, for every $j=1,\ldots,k-1$, the
parent of $i_j$ is also parent of the parent of $i_{j+1}$: if we let
$v_j$ be the parent of $i_j$, for every $j=1,\ldots, k-1$, we obtain a
path $(v_1,\ldots,v_k)$ consisting of tree nodes (because each node in
it has out-degree at least 2) and such that each $v_j$ is the parent
of the leaf $i_j$.

Now, $v_k$ may be either intermediate in the reticulation cycle $K$
for $h$ or the split node of $K$ (in which case one of the merge paths
would be the arc $(v_k,h)$).  But, if the latter happened, $h$ would
have another parent $v$ and it would be a descendant of $v_k$, and
then, any tree descendant leaf $l$ of $v$ would be such that
$\ell_N(i_k,l)=1$, which would contradict (4).  This implies that
$v_k$ is intermediate in $K$.

Let now $v$ be the other parent of $h$, and assume that it is
intermediate in the merge path of $K$ not containing $v_k$.  Let $l$
be a tree descendant leaf of $v$.  By Lemma \ref{lem:1n-cicles}.(d),
$l\notin\{i_1,\ldots,i_k\}$.  Then, by (6),
$\ell_N(i_1,l)=\ell_N(i,l)=2$ and $\ell_N(l,i)=\ell_N(l,i_1)$.  But
the latter condition implies that $[l,i_1]=[l,i]=v$, and then the
former implies that $v$ is the parent of $v_1$, which would imply that
$i_1\in C(v)$, leading to a contradiction again by Lemma
\ref{lem:1n-cicles}.(d).  We conclude that the merge path not
containing $v_k$ is a single arc.  In particular, this implies that no
node $v_1,\ldots,v_{k-1}$ is the split node of $K$: if $v_j$ were the
split node of $K$, then $\ell_N(i,i_j)=2$, against (5).  So, the split
node $u$ of $K$ is a proper ancestor of $v_1$.  Let us see that $u$ is
the parent of $v_1$.  Indeed, if $u$ were not the parent $w$ of $v_1$,
then $w$ would be intermediate in the merge path $u\pathgr v_1\pathgr
h$: let $w'$ be a child of $w$ outside $K$, and let $l$ be a tree
descendant leaf of $w'$.  Then, since $l\notin\{i,i_1,\ldots,i_k\}$,
(6) would imply that $\ell_N(i,l)=\ell_N(i_1,l)=2$, while it is clear
that $\ell_N(i,l)=k+2$ (because $[l,i]=w$ and the only path $w\pathgr i$,
along the merge path, has length $k+2$).

In summary, we have proved so far that if $N$ satisfies
conditions (1) to (6), then it contains a reticulation cycle for the
hybrid parent $h$ of $i$ consisting of the merge paths
$(u,v_1,\ldots,v_k,h)$ and $(u,h)$, and that each $v_j$ is the parent
of the tree leaf $i_j$.  It remains to prove that $v_1,...,v_k$ have
out-degree 2.  But, if some $v_j$ had some child $w_j$ other than $i_j$
or its child in $K$, and if $l$ were a tree descendant leaf of $w_j$,
then $l\notin\{i,i_j,\ldots, i_k\}$ but $\ell_N(i_j,l)=1$, against
(4).
\end{proof}


\begin{lemma}\label{lem:nod-rG2}
The $G_{i;i_1,\ldots,i_k;i'_1,\ldots,i'_{k'}}$ reduction (with $k\geq
k'> 0$) can be applied to $N$ if, and only if, the following
conditions are satisfied:
\begin{itemize}
\item[(1)] $\ell_N(i,l)>1$ for every $l\in S\setminus\{i\}$.

\item[(2)] $\ell_N(i_k,i)=1$ and $\ell_N(i,i_k)=2$.

\item[(2')] $\ell_N(i'_{k'},i)=1$ and $\ell_N(i,i'_{k'})=2$.

\item[(3)] For every $j=1,\ldots,k-1$, $\ell_N(i_j,i_{j+1})=1$ and
$\ell_N(i_{j+1},i_j)=2$.

\item[(3')] For every $j=1,\ldots,k'-1$, $\ell_N(i'_j,i'_{j+1})=1$ and
$\ell_N(i'_{j+1},i'_j)=2$.

\item[(4)] For every $j=1,\ldots,k$ and for every $l\neq
i_{j},...,i_k,i$, $\ell_N(i_j,l)>1$.

\item[(4')] For every $j=1,\ldots,k'$ and for every $l\neq
i'_{j},...,i'_{k'},i$, $\ell_N(i'_j,l)>1$.

\item[(5)] $\ell_N(i_1,i'_1)=\ell_N(i'_1,i_1)=2$.
%


%

\end{itemize}
\end{lemma}
 
\begin{proof}
Assume that $N$ contains a reticulation cycle $K$ consisting of the
merge paths $(u,v_1,\ldots,v_k,h)$ and $(u,v'_1,\ldots,v'_{k'},h)$,
with $k\geq k'> 0$, such that the only child of the hybrid node $h$ is
the leaf $i$, each tree node $v_j$ has only one child outside $K$, and
it is the tree leaf $i_j$, and each tree node $v'_j$ has only one
child outside $K$, and it is the tree leaf $i'_j$.  The proof that it
satisfies the conditions (1) to (4) and (1') to (4') is similar to the
corresponding proof in the previous lemma, and (5) is a direct
consequence of the fact that $[i_1,i'_1]=u$ (because $v_1$ and $v_1'$
are not connected by a path by Lemma \ref{lem:1n-cicles}.(a)) .

Conversely, assume that $N$ satisfies all conditions listed in the
statement.  Conditions (1), (2) and (2') imply that $i$ is a
quasi-sibling of $i_k$ and $i'_{k'}$: let $h$ be the hybrid parent of
$i$, and let $v_k$ and $v'_{k'}$ be, respectively, the parents of
$i_k$ and $i'_{k'}$.  As in the previous lemma, conditions (3) and
(3') imply the existence of paths $(v_1,\ldots,v_k)$ and
$(v'_1,\ldots,v'_{k'})$ consisting of tree nodes and such that each
$v_j$ is the parent of the leaf $i_j$ and each $v'_j$ is the parent of
the leaf $i'_j$.

Now, no node $v_1,\ldots,v_k,v'_1,\ldots,v_{k'}$ is the split node of
$K$: if, say, $v_j$ were the split node of $K$, then in particular
$v'_{k'}$, and hence $i'_{k'}$, would be a descendant of $v_j$, which
would imply that $v_j=[i_j,i'_{k'}]$ and thus
$\ell_{N}(i_j,i'_{k'})=1$, against (4).  Therefore, the split node of
$K$ is a common ancestor of $v_1$ and $v_1'$.  Now, (6) implies that
$[i_1,i_1']$ is simultaneously the parent of $v_1$ and $v_1'$, and
therefore that this parent is the split node $u$ of $K$.

Finally, the proof that the intermediate nodes of $K$ have out-degree
2 is similar to the proof of the corresponding fact in the previous
lemma, using (4) and (4').
\end{proof}

Lemmas \ref{lem:nod-rR} to \ref{lem:nod-rG2} imply that the
possibility of applying a specific $R$, $T$ or $G$ reduction to $N$
depends only on $\ell(N)$, from where condition (A) in Theorem
\ref{prop:snod-AR} follows.  As far as condition (R) goes, we have the
following lemma.

\begin{lemma}\label{lem:nod-R}
\begin{enumerate}[(a)]
\item If the $R_{i;j}$ reduction can be applied to $N$, then, for
every $k,l\in S\setminus\{j\}$,
  \begin{itemize}
  \item $\ell_{R_{i;j}(N)}(i,k)=\ell_N(i,k)-1$ if $k\neq i$.
  \item $\ell_{R_{i;j}(N)}(k,i)=\ell_N(k,i)$  if $k\neq i$.
  \item $\ell_{R_{i;j}(N)}(k,l)=\ell_N(k,l)$  if $k,l\neq i$.
  \end{itemize}

\item If the $T_{i;j}$ reduction can be applied to $N$, then, for
every $k,l\in S\setminus\{j\}$,
$$
\ell_{T_{i;j}(N)}(k,l)=\ell_N(k,l).
$$

\item If the $G_{i;i_1,...,i_k,i'_1,...,i'_{k'}}$ reduction (with $k\geq k'\geq 0$)  can be applied to $N$, then, for every $j,l\in S\setminus\{i_2,\ldots,i_k,i'_1,\ldots,i'_{k'}\}$,
  \begin{itemize}
  \item $\ell_{G_{i;i_1,...,i_k,i'_1,...,i'_{k'}}(N)}(i,j)=\ell_N(i_1,j)-1$ if $j\neq i_1,i$
  \item $\ell_{G_{i;i_1,...,i_k,i'_1,...,i'_{k'}}(N)}(j,i)=\ell_N(j,i)$  if $j\neq i_1,i$
   \item $\ell_{G_{i;i_1,...,i_k,i'_1,...,i'_{k'}}(N)}(i_1,j)=\ell_N(i_1,j)-1$ if $j\neq i_1,i$
  \item $\ell_{G_{i;i_1,...,i_k,i'_1,...,i'_{k'}}(N)}(j,i_1)=\ell_N(j,i_1)$  if $j\neq i_1,i$ 
  \item $\ell_{G_{i;i_1,...,i_k,i'_1,...,i'_{k'}}(N)}(i,i_1)=\ell_N(i_1,i)=1$ 
  \item $\ell_{G_{i;i_1,...,i_k,i'_1,...,i'_{k'}}(N)}(j,l)=\ell_N(j,l)$  if $j,l\neq i_1,i$ 
  \end{itemize}
\end{enumerate}
\end{lemma}

\begin{proof}
\emph{(a)} $R_{i;j}(N)$ is obtained by removing the leaf $j$ and
replacing the leaf $i$ by its parent.  This implies that, for every
pair of remaining leaves, their LCA is the same node in $N$ and in
$R_{i;j}(N)$, and that any path ending in $i$ is shortened in one arc,
while all paths ending in any other remaining leaf are left untouched.
The formulas for $\ell_{R_{i;j}(N)}$ given in the statement follow
immediately from these observations.  \smallskip

\noindent\emph{(b)} $T_{i;j}(N)$ is obtained by removing the leaf $j$
without modifying anything else.  This implies that, for every pair of
remaining leaves, their LCA is the same node in $N$ and in
$T_{i;j}(N)$ and no path ending in a remaining leaf is modified, and
therefore that $\ell_{T_{i;j}(N)}=\ell_N$ on $S\setminus\{j\}$.
\smallskip

\noindent\emph{(c)} Let us denote by $N'$ the network
$G_{i;i_1,...,i_k,i'_1,...,i'_{k'}}(N)$, and let $u$ be the split node
of the removed reticulation cycle.  We remove all (and only)
descendants of $u$, and we add to $u$ two new tree leaf children $i$
and $i_1$.  This implies that the LCA in $N'$ of $i$ and $i_1$ is $u$
(and therefore $\ell_{N'}(i,i_1)=\ell_{N'}(i_1,i)=1$) and that the LCA
of any other pair of remaining leaves is the same node in $N'$ as in
$N$.  On the other hand, any path ending in $i_1$ is shortened in one
arc, the distance from any internal node to $i$ in $N'$ is the same as
its distance to $i_1$, and all paths ending in remaining leaves other
than $i$ or $i_1$ are not touched.  From these observations, the
formulas for $\ell_{N'}$ given in the statement easily follow.
\end{proof}

\end{document}